\documentclass[twocolumn,showpacs,showkeys]{revtex4}
\usepackage{graphicx}
\usepackage{subfigure}
\usepackage{bm}

\def\vec#1{\mathchoice{\mbox{\boldmath$\displaystyle#1$}}
{\mbox{\boldmath$\textstyle#1$}}
{\mbox{\boldmath$\scriptstyle#1$}}
{\mbox{\boldmath$\scriptscriptstyle#1$}}}



\begin{document}

\title{Rapidity-dependent spectra from a single-freeze-out 
model of relativistic heavy-ion collisions}
\author{Bart\l{}omiej Biedro\`n}
\email{rockhouse@dione.ifj.edu.pl}
\affiliation{AGH University of Science and Technology, Al. Mickiewicza 30, 30-059
Krak\'ow, Poland}
\author{Wojciech Broniowski}
\email{Wojciech.Broniowski@ifj.edu.pl}
\affiliation{Institute of Physics, \'Swi\c{e}tokrzyska Academy,
ul.~\'Swi\c{e}tokrzyska 15, PL-25406~Kielce, Poland} 
\affiliation{The H. Niewodnicza\'nski Institute of Nuclear Physics, Polish Academy of
Sciences, PL-31342 Krak\'ow, Poland}

\date{ver.~2, 20 November 2006}

\begin{abstract}
An extension of the single-freeze-out model with thermal and geometric
parameters dependent on the
spatial rapidity, $\alpha_\parallel$, is used to describe the rapidity 
and transverse-momentum
spectra of pions, kaons, protons, and antiprotons measured at RHIC 
at $\sqrt{s_{NN}}=200~{\rm GeV}$ by the
BRAHMS collaboration. {\tt THERMINATOR} is used to perform the
necessary simulation, which includes all resonance decays. The result of the fit to 
the rapidity spectra 
in the range of the BRAHMS data is the expected growth of the baryon and strange
chemical potentials with the magnitude of $\alpha_\parallel$, while the 
freeze-out temperature is kept fixed. The value of the baryon chemical potential 
at $\alpha_\parallel \sim 3$, which is the relevant region for 
particles detected at the BRAHMS forward rapidity $y \sim 3$, is about $200~{\rm GeV}$, 
{\em i.e.} lies in the 
range of the values obtained for the highest SPS energy. 
The chosen geometry of the fireball
has a decreasing transverse size as the magnitude of  $\alpha_\parallel$ 
is increased, which also corresponds to decreasing transverse flow. This feature is 
verified by reproducing the 
transverse momentum spectra of pions and kaons at various rapidities. 
The strange chemical potential obtained from the fit to the $K^+/K^-$ ratio is such that
the local strangeness density in the fireball is compatible with zero.
The resulting rapidity spectra of net protons 
are described qualitatively in the model.  
As a result of the study, the knowledge of the ``topography'' of the fireball is achieved, 
making other calculations possible. 
As an example, we give predictions for the rapidity spectra of hyperons.
\end{abstract}

\maketitle

\section{Introduction \label{intro}}

The study of particle abundances has been a major source of information
concerning heavy-ion collisions. In fact, the agreement of the particle ratios
with simple predictions of statistical models 
%\cite{Fermi:1950jd,Pomeranchuk:1951ey,Landau:1953gs,Belenkij:1956cd} 
is a key argument for early thermalization of the formed system
\cite{br:w,phob:w,st:w,phen:w}. Up to now the
numerous studies of the particle ratios \cite%
{Koch:1985hk,Cleymans:1992zc,Sollfrank:1993wn,Braun-Munzinger:1994xr,%
Braun-Munzinger:1995bp,Gazdzicki:1998vd,Yen:1998pa,Cleymans:1998fq,%
Becattini:2000jw,Rafelski:2000by,Braun-Munzinger:2001ip,Michalec:2001um,%
Magestro:2001jz,Becattini:2003wp,Braun-Munzinger:2003zd,Torrieri:2003nh,%
Cleymans:2004pp,Rafelski:2004dp,Becattini:2005pt}
were falling into two basic categories: the so-called $4\pi$ studies at low
energies (SIS, AGS) and the studies at mid-rapidity for approximately
boost-invariant systems at highest energies (RHIC). The $4\pi$ studies
involve three-momentum integrals of the statistical distribution functions,
with the multiplicity of species $i$ given by $N_{i}=V \int d^{3}p f_{i}( 
\sqrt{m_{i}^{2}+p^{2}})$, thus providing information on
volume-averaged thermal parameters of the system in a very  
simple way. The inclusion of resonance decays \cite%
{Brown:1991en,Sollfrank:1990qz,Bolz:1992hc}, crucial for the success of the
approach, is also straightforward, since the detection of the products with
full angular coverage is insensitive to the decay kinematics or flow
effects; once the resonance has decayed, its products are 
registered. The other simple situation arises when the system is nearly
boost-invariant. To a sufficiently good accuracy this is the case at 
mid-rapidity for the RHIC energies, where the particle yields $dN/dy$ 
change only by a few percent 
in the rapidity window $|y| < 1$. The assumption of the boost-invariance of the fireball 
leads again to very simple formulas. 
Although the particles detected at mid-rapidity are
collected from various parts of the fireball, not only from the very central
region, their ratios are the same as in the $4\pi$ calculation. This is
because for $dN_{i}/dy=\int d^{2}p_{\perp }d^3N_{i}/(d^{2}p_{\perp }dy)$ we
have from the boost invariance \cite{Broniowski:2001we} 
\begin{equation}
\frac{dN_{i}/dy}{dN_{j}/dy}=\frac{\int dy\,dN_{i}/dy}{\int dy\,dN_{j}/dy}=%
\frac{N_{i}}{N_{j}}.  \label{boost}
\end{equation}
This obvious general formula finds an explicit manifestation in specific
boost-invariant models. The result also holds when resonance decays are
included, see Ref.~\cite{Broniowski:2002nf} for a derivation in the framework of
the Cooper-Frye \cite{Cooper:1974mv} formalism.

\begin{figure}[tb]
\begin{center}
\includegraphics[width=.47\textwidth]{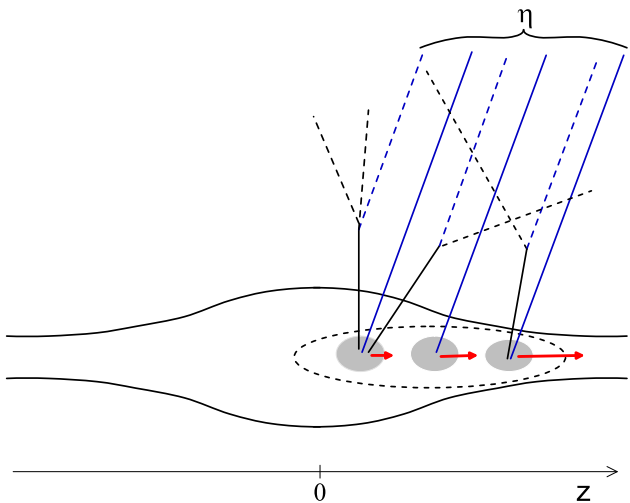}
\end{center}
\vspace{1mm}
\caption{(Color online) Emission of particles from a boost-non-invariant
fireball. The horizontal (vertical) axis indicates longitudinal coordinate $z$ 
(transverse coordinate $\rho$).
Particles emitted with the same value of pseudorapidity $\eta$
originate from different regions of the fireball indicated by the 
gray blobs (they also are emitted at different times). The thermal
conditions and flow (indicated by arrows) in these regions are different. 
The solid lines
indicate tracks of primordial particles, while the dashed lines show products of
resonance decays. The dashed ellipse indicates the relevant region for a give $\eta$, 
which spans about two units of the spatial rapidity $\alpha_\parallel$.}
\label{fig:fireball}
\end{figure}

When the system is not boost invariant the above simplifications no longer
hold. The situation is illustrated in Fig.~\ref{fig:fireball}. Particles
detected at a given pseudorapidity $\eta$ (parallel lines in the figure) originate
from different pieces of the fireball (gray blobs). Thermal conditions
(temperature, chemical potentials, flow) change from piece to piece, which must be
properly included. In addition, the effects of the longitudinal flow
(indicated by arrows) must be incorporated, and the kinematics of resonance
decays (dashed lines) becomes relevant. Although the resulting formalism for
particle spectra remains conceptually simple and is based on the standard
Cooper-Frye treatment, the calculation is no longer semi-analytic and a
full-fledged simulation is necessary to accomplish the goal. 

In the analysis of this paper we use {\tt THERMINATOR} -- the THERMal heavy IoN generATOR 
\cite{Kisiel:2005hn}, to generate the Monte Carlo events in a suitably modified
single-freeze-out model of Ref.~\cite{Broniowski:2001we}.
The extension to the boost-non-invariant
case consists of two basic elements. The first one (geometric) is the choice of the
shape of the freeze-out hypersurface $\Sigma$ and collective expansion. The
other one incorporates the dependence of the thermal parameters on the
position within the hypersurface $\Sigma$. 
Specifically, in our treatment the transverse size and the chemical potentials depend 
on the  spatial rapidity $\alpha_\parallel=\mathrm{arctanh}(z/t)$, where $z$ and $t$ 
are the longitudinal and time coordinates on the freeze-out hypersurface. 
Although the boost-non-invariant 
model has quite a few parameters, as
listed at the end of Sect.~\ref{model}, they can be fitted
independently to various combinations of the data, leaving little freedom. 
For instance, the $\alpha_\parallel$ dependence of the baryon and strange 
chemical potentials is fixed with the ratios of 
protons to antiprotons and $K^+$ to $K^-$. 
The result in the range of the BRAHMS data is the expected growth of 
the baryon and strange
chemical potentials with $|\alpha_\parallel|$. The value of the baryon 
chemical potential 
$\mu_B$ at $\alpha_\parallel \sim 3$, which is the relevant region for 
particles detected at the BRAHMS forward rapidity, $y \sim 3$, is about $200~{\rm GeV}$. 
This value is in the 
range of the values of the thermal fits for the highest SPS energy, 
thus we confirm the recent findings by Roehrich
\cite{Roehrich} that the thermal conditions at RHIC at forward rapidities, $y \sim 3$,
correspond to the SPS conditions at mid-rapidity. Details of our procedure 
of determining the dependence of thermal parameters on $\alpha_\parallel$ are explained 
in Sect.~\ref{sec:results}. Our strategy, 
as usual, is to fix the features of the fireball with the well-measured spectra 
of particles: pions, kaons, protons and antiprotons. 
The strange chemical potential obtained from the fit to the $K^+/K^-$ ratio is such that
the local strangeness density on the freeze-out hypersurface $\Sigma$ 
is compatible with zero.
The experimental \cite{Bearden:2003fw} rapidity spectra of net protons, $p - \bar p$,
are reproduced qualitatively in the model, displaying the 
correct shape but
overshooting the data at larger rapidities by about 50\%.  
The chosen geometry of the fireball incorporates a decreasing transverse size as 
$|\alpha_\parallel|$ is increased, which simultaneously
results in a decreasing transverse flow. This choice is verified in 
Sect.~\ref{sec:results}  by reproducing the 
spectra $d^2N/(2\pi p_T dp_T dy)$ at a fixed $y$ 
of pions and kaons at 
$\sqrt{s_{NN}}=200~\mathrm{GeV}$ from the BRAHMS
collaboration \cite{Bearden:2004yx}.
These transverse-momentum spectra exhibit slopes which become steeper with rapidity. 
 
As a result of our study, we obtain the ``topography'' of the fireball, which can be
the ground for other more detailed studies, discussed in the Conclusion.

\section{The single freeze-out model \label{model}}

The single-freeze-out model is described in detail in Refs.~\cite%
{Broniowski:2001we,Broniowski:2001uk,Broniowski:2002nf}. Here we review
the main assumptions and the formalism of describing the 
expansion and particle decays.

\begin{enumerate}
\item At a certain stage of evolution of the fireball the thermal
equilibrium between hadrons is reached. Most probably, hadrons are ``born''
already in such an equilibrated state. The local particle phase-space
densities have the form of the Fermi-Dirac or Bose-Einstein statistical
distributions. The particles generated at freeze-out are termed \emph{%
primordial}. For simplicity of the model we do not include the $\gamma$
non-equilibrium factors of 
Ref.~\cite{Rafelski:1991rh} used in several recent analyses 
\cite{Torrieri:2003nh,Becattini:2003wp,Cleymans:2004pp}.

\item The thermodynamic parameters are the freeze-out temperature $T$ and three chemical
potentials: baryon, $\mu_B$, strange, $\mu_S$, and $\mu_{I_3}$, related to
the third component of isospin. In a boost-non-invariant model the values of
these parameters depend on the position within the freeze-out hypersurface $\Sigma$.

\item In the boost-non-invariant model the shape of the fireball is
nontrivial in the longitudinal direction. In this paper we retain the
azimuthal symmetry.

\item The velocity field of the collective expansion is chosen in the form of 
the Hubble flow \cite{Csorgo:1995bi},
providing the longitudinal and transverse flow to the system. Again, in the
boost-non-invariant model the functional form of the velocity field may depend
on the longitudinal position.

\item The evolution after freeze-out includes decays of resonances which
may proceed in cascades. All resonances from the Particle Data Tables \cite%
{Hagiwara:2002fs} are incorporated.

\item Elastic rescattering among particles after the chemical freeze-out is
ignored and the model may be viewed as an approximation to a more detailed
evolution, taking into account different time scales for various hadronic
processes (see \cite{Heinz} and references therein).
\end{enumerate}

The single freeze-out concept complies to the \emph{explosive scenario} at
RHIC \cite{Rafelski:2000by}. Moreover, the approach reproduces very
efficiently the particle abundances, the transverse-momentum spectra,
including particles with strangeness \cite{Broniowski:2001uk}, produces very
reasonable results for the resonance production \cite{Broniowski:2003ax},
the charge balance functions in rapidity \cite{Bozek:2003qi}, the elliptic flow 
\cite{Broniowski:2002wp}, the HBT radii \cite{Kisiel:2006is}, and the transverse
energy \cite{Prorok:2004wi,Prorok:2005jf,Prorok:2005uv}. 
Recently with the help of RQMD \cite{Sorge:1989vt}
it was found (cf.~Fig.~16 and 17 of Ref.~\cite{Nonaka:2006yn}) that 
the elastic rescattering effects are not
significant for the mid-rapidity $p_T$ spectra of pions. One can understand
this as follows: resonance decays ``cool'' the spectra \cite%
{Broniowski:2001we}. As the result, the original $\sim 165~\mathrm{MeV}$
spectra from the chemical freeze-out look, after feeding from resonances,
approximately as $\sim 130~\mathrm{MeV}$ spectra, which would be obtained at
the the lower temperature of the thermal freeze-out. Thus elastic 
rescattering becomes innocuous. 
Certainly, more studies are
desirable here, in particular an ``afterburner'' performing
elastic rescattering could be run
on top of our simulation. It would help to achieve a more accurate collision
picture, with the elastic rescattering processes taken fully into account.

Popular choices of the freeze-out hypersurface and the collective velocity
field are discussed in detail in Ref.~\cite{Florkowski:2004tn}. In this work
we modify in a very simple way 
the original boost-invariant single-freeze-out model of Ref.~\cite%
{Broniowski:2001we,Broniowski:2001uk,Broniowski:2002nf} by making the
transverse size of the fireball dependent on the spatial rapidity. We
use the freeze-out hypersurface parameterized as
\begin{eqnarray}
x^\mu = \left ( 
\begin{array}{c}
t \\ 
x \\ 
y \\ 
z%
\end{array}
\right ) = \left( 
\begin{array}{l}
\tau \mathrm{cosh} \alpha_\perp \mathrm{cosh } \alpha_\parallel \\ 
\tau \mathrm{sinh} \alpha_\perp \cos\phi \\ 
\tau \mathrm{sinh} \alpha_\perp \sin\phi \\ 
\tau \mathrm{cosh} \alpha_\perp \mathrm{sinh } \alpha_\parallel%
\end{array}
\right).  \label{x}
\end{eqnarray}
The parameter $\alpha_\parallel$ is the \emph{spatial rapidity}, while $%
\alpha_\perp$ is related to the transverse radius as 
\begin{eqnarray}
\rho = \sqrt{x^2+y^2} = \tau \mathrm{sinh} \alpha_\perp.  \label{eq:rho}
\end{eqnarray}
The four-velocity field is chosen to follow the Hubble law%
\begin{equation}
u^\mu = x^\mu/\tau.  \label{u}
\end{equation}
We note that the longitudinal flow is $v_z = \hbox{tanh}\, \alpha_\parallel
= z/t$ (as in the one-dimensional Bjorken model \cite{Bjorken:1983qr}),
while the transverse flow (at $z=0$) has the form $v_\rho = \hbox{tanh}\,\alpha_\perp$.

The new element of the parameterization of this paper, which implements the
departure from boost invariance, is the selection of boundaries for the
fireball. In the boost invariant model $\rho$ was limited by the
space-independent parameter $\rho_{\mathrm{max}}$, or $0 \le \alpha_\perp
\le \alpha_\perp^{\mathrm{max}}$. Now we take 
\begin{eqnarray}
0 \le \alpha_\perp \le \alpha_\perp^{\mathrm{max}}(\alpha_\parallel) \equiv
\alpha_\perp^{\mathrm{max}}(0) \exp \left ( -\frac{\alpha_\parallel^2}{2
\Delta^2}\right ).  \label{eq:Delta}
\end{eqnarray}
The interpretation of this formula is clear: as we depart from the center by
increasing $|\alpha_\parallel|$, we simultaneously reduce $\alpha_\perp$, or 
$\rho_{\mathrm{max}}$. The rate of this reduction is controlled by a new
model parameter, $\Delta$. Since in our model the flow is linked to the
position via Eq.~(\ref{u}), we also have less transverse flow as we increase 
$|\alpha_\parallel|$. This feature will show in the $p_T$ spectra presented
in Sect.~\ref{sec:results}. We may also use more conveniently the parameter
\begin{eqnarray}
\rho_{\mathrm{max}}^{(0)}= \mathrm{sinh}\alpha_\perp^{\mathrm{max}}(0).
\label{rho0}
\end{eqnarray}
Thus the geometry and expansion of the fireball is described by three
parameters: $\tau$, $\rho_{\mathrm{max}}^{(0)}$, and $\Delta$.

With the standard parameterization of the particle four-momentum in terms of
rapidity $y$ and the transverse mass $m_\perp$, 
\begin{equation}
p^\mu = \left(m_\perp \hbox{cosh} y, p_\perp \cos\varphi, p_\perp
\sin\varphi, m_\perp \hbox{sinh} y \right),  \label{pmubinv}
\end{equation}
we find with Eqs.~(\ref{x}) and (\ref{u}) 
\begin{equation}
p \cdot u = m_\perp \hbox{cosh}(\alpha_\perp) \hbox{cosh}(\alpha_%
\parallel-y) - p_\perp \hbox{sinh}(\alpha_\perp) \cos(\phi-\varphi),
\label{eq:pu}
\end{equation}
and 
\begin{eqnarray}
d^3\Sigma \cdot p \!\!&=& \!\! d\alpha_\parallel d\phi \, \rho \, d\rho
\times  \nonumber \\
&& \left [ m_\perp \sqrt{\tau^2+\rho^2} \, \hbox{cosh}(\alpha_\parallel-y) -
p_\perp \rho \cos(\phi-\varphi) \right ]  \nonumber \\
&=& \tau^3 d\alpha_\parallel d\phi \,\mathrm{sinh}\alpha_\perp \,
d\alpha_\perp \, p \cdot u  \label{sigmapbinv}
\end{eqnarray}
where $d^3\Sigma^\mu$ is the volume element of the hypersurface. With the assumed
azimuthal symmetry the Cooper-Frye \cite{Cooper:1974mv} formalism then
yields the following expression for the momentum density of a given species
of primordial particles: 
\begin{eqnarray}
\frac{d^2N}{ 2\pi p_T dp_T \,dy} &=& \tau^3 \!\! \int_{-\infty}^\infty
\!\!\!\!\!\! d\alpha_\parallel \int_0^{\alpha_\perp^{\mathrm{max}%
}(\alpha_\parallel)}\!\!\!\!\!\!\!\!\!\!\!\!\! d\alpha_\perp \int_0^{2\pi}
\!\!\!\!\!\! d\phi  \nonumber \\
&&\times p \cdot u f\left ( \beta p \cdot u - \beta
\mu(\alpha_\parallel)\right ),  \label{master} \\
f(z)&=&\frac{1}{(2\pi)^3} \frac{1}{\exp z \pm 1 },  \nonumber  \label{mod}
\end{eqnarray}
where $p \cdot u$ from Eq.~(\ref{eq:pu}) is taken at $\varphi = 0$, $f(z)$
is the statistical distribution function (with $+$ for fermions and $-$ for
bosons), $\beta=1/T$, and 
\begin{eqnarray}
\mu(\alpha_\parallel) =B \mu_B(\alpha_\parallel) + S \mu_S(\alpha_\parallel)
+I_3 \mu_{I_3}(\alpha_\parallel),  \label{eq:mu}
\end{eqnarray}
with $B$, $S$, and $I_3$ denoting the baryon number, strangeness, and the
third component of isospin of the particle. Thus we admit the dependence of
chemical potentials on the spatial rapidity. This is of course necessary if
we wish to describe in the framework of a statistical model the increasing
density of net protons as we move from mid-rapidity towards the fragmentation region.

The temperature $T$ also in general depends on $\alpha_\parallel$. The
best model-building strategy here would be to use the universal 
Cleymans-Redlich chemical freeze-out
curve  \cite{Cleymans:1998fq} in the $\mu_B$-$T$ space (for a recent status see 
Ref.~\cite{Becattini:2003wp,Cleymans:2006qe}). That way
the functional dependence of $\mu_B$ on $\alpha_\parallel$ induces
unambiguously the dependence of $T$ on $\alpha_\parallel$. In this work, however, we
apply the model for not too large values of the rapidity, $|y| \le 3.3$, and it will
turn out that the obtained values for $\mu_B$ are less than $\sim 250~%
\mathrm{MeV}$. The universal freeze-out curve gives from $\mu_B=0$ to $%
\mu_B=250~\mathrm{MeV}$ a practically constant value of $T$. For instance,
at SPS ($Pb+Pb$, $\sqrt{s_{NN}}=17~\mathrm{GeV}$) one has \cite%
{Michalec:2001um} $T=164 \mathrm{~MeV}$, $\mu_B=229\mathrm{~MeV}$, $\mu_S=54%
\mathrm{~MeV}$, and $\mu_{I_3}=-7\mathrm{~MeV}$, with the value of $T$ equal
within errors to the RHIC value of 165~MeV. For this reason in the present
analysis we fix the freeze-out temperature at a constant (independent of 
the spatial rapidity) value, 
\begin{equation}
T=165~\mathrm{MeV}.  \label{T}
\end{equation}
If modeling were made for larger values of the rapidity and/or lower
collision energies, the dependence of $T$ on $\alpha_\parallel$ should be
incorporated according to the prescription based on the universal freeze-out curve. 
Eventually, we expect that
when the fragmentation region is approached, $T$ becomes very small and $%
\mu_B$ reaches the value of the order of 1~GeV.

Another qualitative argument for the approximate constancy 
of $T$ at moderate values of $|\alpha_\parallel|$ 
may be inferred directly from the BRAHMS data. From the measured rapidity spectra
(cf. Fig.~\ref{fig:bry}), obviously, the yields of pions and kaons decrease with $y$.
Thus, one needs to decrease the size of the emitting source, decrease the 
freeze-out temperature, 
or both. The decrease of 
temperature affects more strongly the particles with larger
masses, since the thermal factor is approximately $\exp(-\sqrt{m^2+{\vec{p}}^2}/T)$. 
Therefore, if we introduced variation of the temperature with the spatial rapidity, it would 
result in a faster drop with $y$ of the pion yields compared to the kaons. 
Certainly the data excludes 
this situation, since the ratio of $dN_\pi/dy$ to $dN_K/dy$ is within a few percent independent of $y$ 
in the BRAHMS rapidity coverage. Therefore we must keep $T$ constant (within a few percent), 
and the remaining possibility is the decrease of the 
source size with $|\alpha_\parallel|$. 
Resonance decays complicate the above qualitative argument, but with 
the help of a numerical simulation
we confirm it.
Another way of providing the drop of yields with rapidity is to incorporate 
the $\gamma$ non-equilibrium factors 
\cite{Rafelski:1991rh} dependent on $\alpha_\parallel$, which may dilute the system 
as $|\alpha_\parallel|$ increases. We do not explore this possibility here.   

\begin{figure}[tb]
\begin{center}
\subfigure{\includegraphics[width=.38\textwidth]{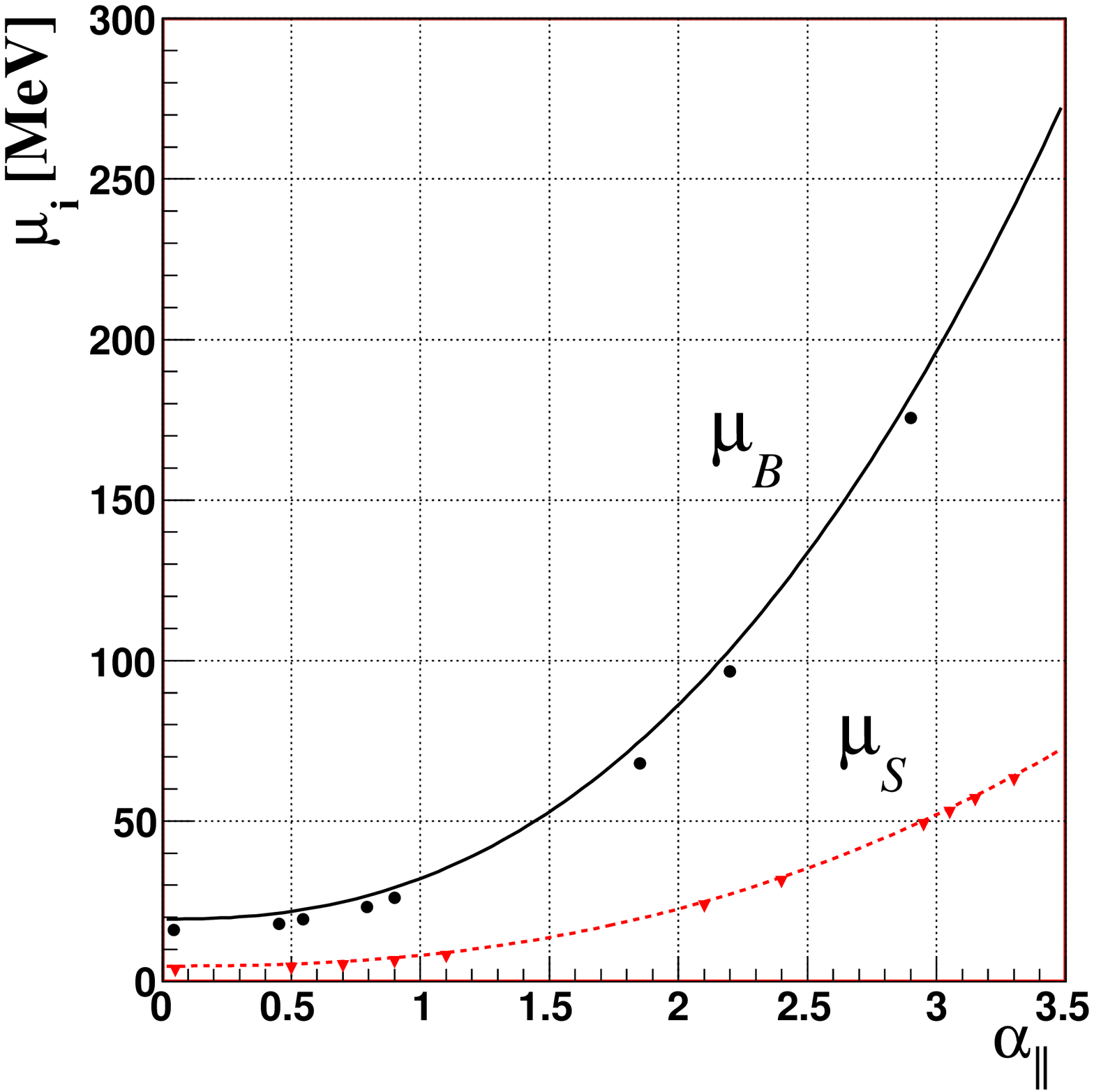}}\\
\vspace{-1mm}
\subfigure{\hspace{1.5mm} \includegraphics[width=.4\textwidth]{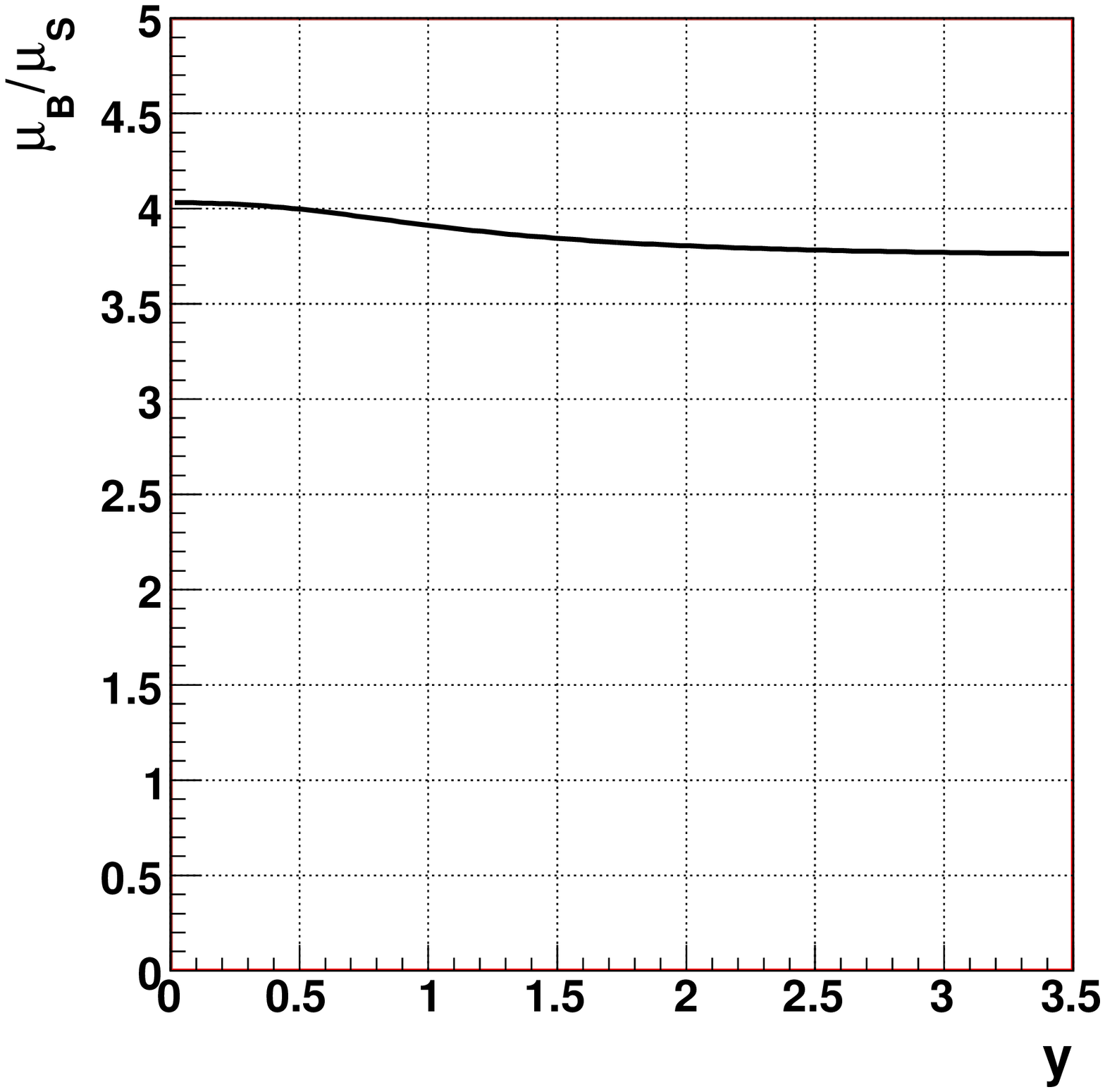}}
\end{center}
\vspace{-9mm}
\caption{(Color online)
Top: the model baryon and strange chemical potentials 
plotted as functions of the spatial
rapidity. Parameters of Eq.~(\ref{ff}) are obtained from the fit to the 
BRAHMS data \protect\cite{Bearden:2003fw,Bearden:2004yx}. The points represent 
a naive calculation based of Eq.~(\ref{nai}). Bottom: the ratio of the baryon to strange
chemical potentials, $\mu_B/\mu_S$. 
\label{fig:muy}}
\end{figure}

For convenience, we parameterize functionally 
the dependence of the chemical potentials at low values of 
$|\alpha _{\parallel }|$ as follows: 
\begin{equation}
\mu _{i}(\alpha _{\parallel })=\mu _{i}(0) \left [1+A_{i}\alpha _{\parallel
}^{2.4} \right],\;\;\; i=B,S,I_{3}. \label{ff}
\end{equation}
The chosen power of 2.4 works somewhat better than 2. Of course, any 
convenient and sufficiently rich parametric form is admissible here, 
as it is fitted to the data
(see Sect. \ref{sec:results}) and the introduced parameters effectively are not free.  
By ``low'' $|\alpha _{\parallel }|$ we mean the values
relevant to the BRAHMS data, covering $|y|\leq 3.3$.

Formula (\ref{mod}) provides the spectra of the primordial particles. The
following evolution of the system consists of free streaming, with
resonances decaying into the daughter particles. We use {\tt THERMINATOR}
to perform the simulation. The code incorporates all the **** and
*** resonances. Following the scheme of {\tt SHARE} \cite%
{Torrieri:2004zz} it excludes all * resonances, and practically all
** resonances listed in the Particle Data Tables \cite%
{Hagiwara:2002fs}. Each resonance decays at the time controlled by its
lifetime, $1/\Gamma$. In the resonance's rest frame the decay at time $t$
occurs with the probability density $\Gamma \exp(-\Gamma t)$. The two-body
or three-body decay channels are incorporated and the values of the
branching ratios are taken from the Particle Data Tables. Heavy resonances
may of course decay in cascades.

Let us summarize the model parameters. We have the universal 
freezeout temperature $T$, three chemical potentials at
mid-rapidity, $\mu_B(0)$, $\mu_S(0)$, $\mu_{I_3}(0)$, three parameters
$A_B$, $A_S$, and $A_{I_3}$ of Eq.~(\ref{ff}) describing the dependence of the 
chemical potentials on the spatial
rapidity, and three geometry/expansion parameters:
the proper time $\tau$, the transverse size at mid-rapidity, $\rho_{\mathrm{%
max}}^{(0)}$, and the parameter $\Delta$ controlling the spatial rapidity
dependence of the transverse size. Except for $T$ taken to have the value (\ref%
{T}) obtained in earlier thermal analyses of particle ratios \cite%
{Broniowski:2001we}, the remaining parameters are fitted to the
BRAHMS data \cite{Bearden:2003fw,Bearden:2004yx} for the double
differential spectra $d^2N/(2\pi p_T dp_T dy)$.

\section{Results \label{sec:results}}

\begin{figure}[tb!]
\begin{center}
\vspace{-1mm}
\subfigure{\includegraphics[width=.38\textwidth]{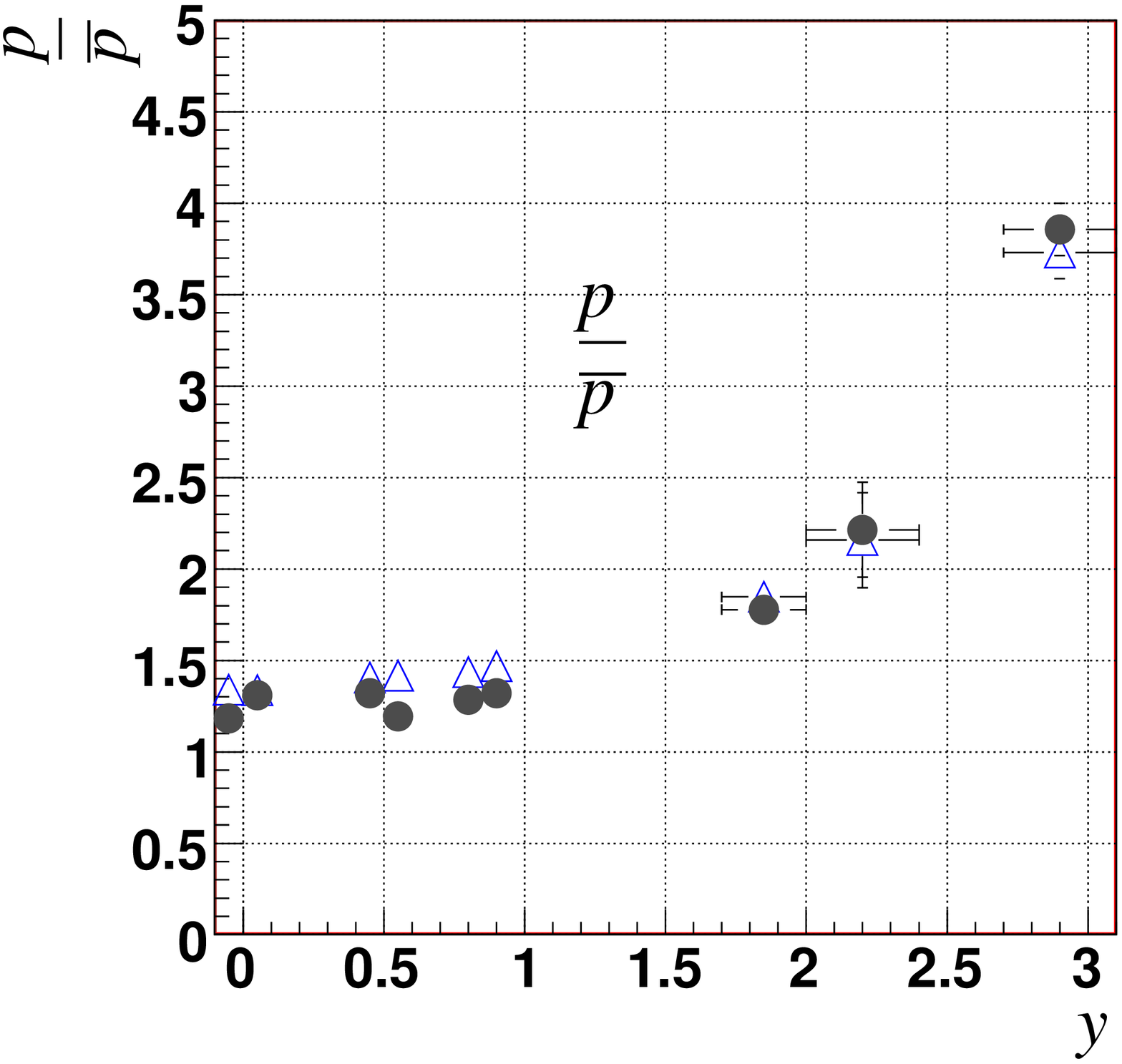}}\\
\vspace{-6mm}
\subfigure{\includegraphics[width=.38\textwidth]{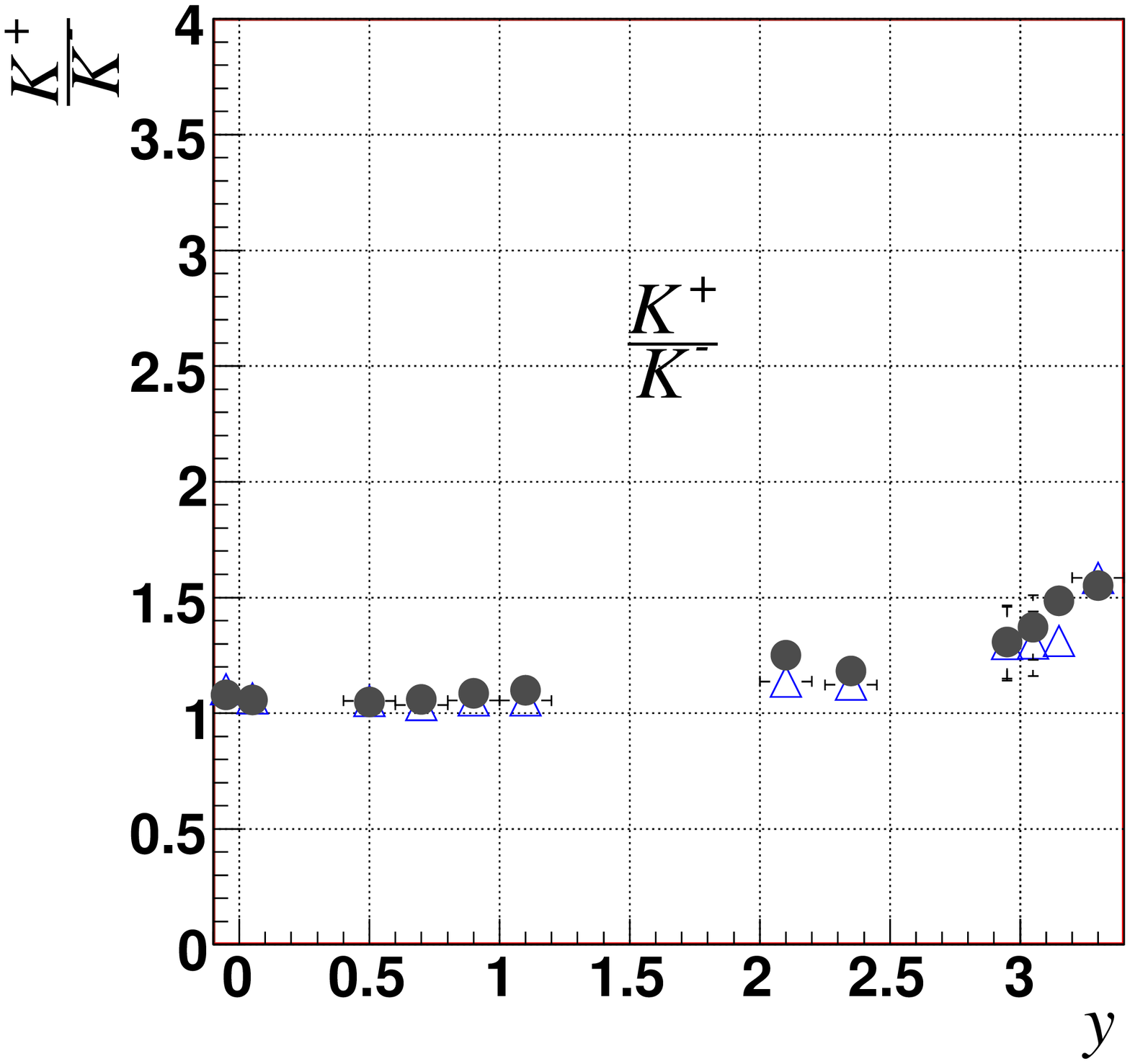}}\\
\vspace{-6mm}
\subfigure{\includegraphics[width=.38\textwidth]{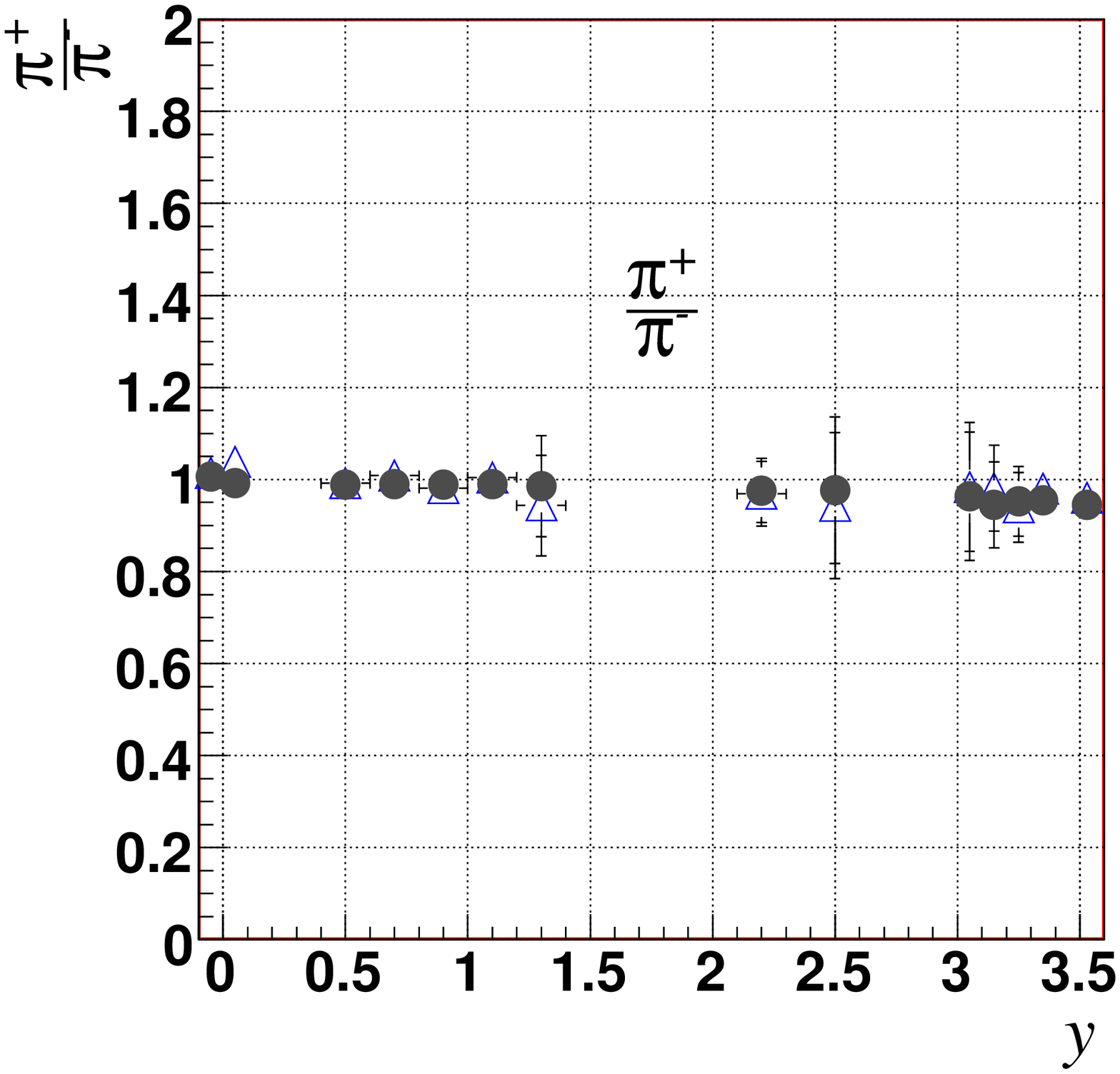}}
\end{center}
\vspace{-9mm}
\caption{(Color online) Dependence of particle ratios on rapidity: top -- $p/%
{\bar p}$, middle -- $K^+/K^-$, bottom -- $\protect\pi^+/\protect\pi^-$.
The open triangles are the Brahms data \protect\cite%
{Bearden:2003fw,Bearden:2004yx}, while the filled circles show the result of the
model simulation with \texttt{THERMINATOR}. The model parameters are 
given in Eq.~(\ref{numpar}). The data for $p$ and $\bar p$ 
are not corrected for the feed-down from weak-decays \cite{Bearden:2003fw}.
\label{fig:ratiosy}}
\end{figure}

We first describe our fitting strategy, which with many parameters present
must be done with care. We wish to have a good starting point for the
parameters describing the chemical potentials. Practice shows that to a
very good approximation the statistical distributions are very well
approximated by the Boltzmann factors. Then the integrand of Eq.~(\ref%
{master}) contains the factor $\exp[-\beta m_\perp \mathrm{cosh}%
(\alpha_\perp) \mathrm{sinh}(\alpha_\parallel - y) + \beta
\mu(\alpha_\parallel)]$. Because of this the relevant integration range in $%
\alpha_\parallel$ is sharply peaked around $\alpha_\parallel \simeq y$ (the
half-width is about one unit of $\alpha_\parallel$) and the chemical
potentials entering the formula are taken approximately at $%
\mu_i(\alpha_\parallel) \simeq \mu_i(y)$. Thus the factors $\exp[\beta \mu(y)%
]$ can be taken in front of the integration. If it were not for the
resonance decays which modify the result (and are included in the full
simulation) we would have to a good approximation the following relations: 
\begin{eqnarray}
\frac{p}{\bar p}&\simeq&\exp (2\beta\mu_B),  \nonumber \\
\frac{K^+}{K^-}&\simeq&\exp (2\beta\mu_S), \;\;\; \mathrm{%
(approximate~formulas)}  \nonumber \\
\frac{\pi^+}{\pi^-}&\simeq&\exp (2\beta\mu_{I_3}).  \label{start}
\end{eqnarray}
The symbols on the left-hand side denote the ratios of yields of the
specified particles at a fixed $y$ and integrated over $p_T$. These are known
form the data, thus we can invert 
\begin{equation}
\mu_B(y)=\frac{1}{2} T \log (p/{\bar p}),\;\;\; \mathrm{%
(approximate~formula)}   \label{nai}
\end{equation}
and so on. With the help of this form we set the starting values of the
parameters $\mu_i(0)$ and $A_i$, which are then iterated. 
The iteration proceeds as follows: 
for a given set of parameters we run the full {\tt THERMINATOR}
simulation, which generates events. We first optimize the baryon-number
parameters $\mu_B(0)$ and $A_B$ with the help of the ratio of the $p$ and ${%
\bar p}$ rapidity spectra, then the strangeness parameters $\mu_S(0)$ and $%
A_S $ using the $K^+$ to $K^-$ ratio, then we go back again to the baryon
parameters, etc., and loop until a fixed is reached. The isospin
parameters $\mu_{I_3}(0)$ and $A_{I_3}$ are consistent with zero and thus
irrelevant. The $\Delta$ parameter is fixed with the pion rapidity spectra 
$dN_{\pi^\pm}/dy$. The optimum value is 
\begin{equation}
\Delta=3.33.  \label{Delta}
\end{equation}

The result of our optimization for the chemical potentials is shown in 
Fig.~\ref{fig:muy}. The optimum parameters are:
\begin{eqnarray}
&& \!\!\!\!\!\!\!\! \mu _{B}(0)=19~{\rm MeV},
\;\mu _{S}(0)=4.8~{\rm MeV},\;\mu _{I_3}(0)=-1~{\rm MeV},  \nonumber \\
&& \!\!\! A_{B}=0.65,\;A_{S}=0.70,\;A_{I_3}=0. \label{numpar}
\end{eqnarray}%
We observe the expected behavior for the baryon chemical potential, which
increases with $|\alpha _{\parallel }|$. The value at the origin is 19~MeV, 
somewhat lower than the earlier mid-rapidity fits made
in boost-invariant models in 
Refs.~\cite{Broniowski:2001we,Braun-Munzinger:2001ip}, yielding 26~MeV. 
The lower value in our case is well understood. The previous mid-rapidity fits 
include the data in the 
range $|y| \le 1$. This range collects the particles emitted from the fireball at 
$|\alpha_\parallel| \le 2$, hence the value of $\mu_B$ in the 
previous mid-rapidity fits is an average of 
our $\mu_B(\alpha_\parallel)$ over the range, approximately, $|\alpha_\parallel| \le 2$, 
with some weight proportional to the particle abundance. 
This qualitatively explains the effect of 
a lower value of our $\mu_B(0)$ than in the boost-invariant models. 
A similar effect occurs for $\mu_S$.
We do not incorporate corrections for the feed-down from weak decays ({em i.e.} 
all decays are 
included), since this is the policy of Ref.~\cite{Bearden:2003fw} for the
treatment of $p$ and $\bar p$.

We note that at $\alpha _{\parallel }=3$
the value of $\mu_B$ is 200~MeV, more than 10~times larger than
at the origin. This value is comparable to the 
highest-energy SPS fit ($\sqrt{s_{NN}}=17~\mathrm{GeV}$), where $\mu
_{B}\simeq 230~\mathrm{MeV}$. The behavior of the strange chemical
potential is qualitatively similar. It also increases with $|\alpha
_{\parallel }|$, growing form 5~MeV at the origin to 50~MeV at $\alpha _{\parallel }=3$.
The ratio $\mu_B(\alpha_\parallel)/\mu_S(\alpha_\parallel)$ 
is very close to a constant, $\simeq 4-3.5$,
as can be seen in the bottom panel of Fig.~\ref{fig:muy}. 

The points in the top panel of Fig.~\ref{fig:muy} show the result of 
the naive calculation of 
Eq.~(\ref{start},\ref{nai}). We note that these points are very 
close (in particular for 
the strangeness case) to the result of the full-fledged fit of our model. 
This is of practical 
significance, since the application of Eq.~(\ref{start},\ref{nai}) 
involves no effort, while the model calculation incorporating resonance decays, flow, 
{\em etc.}, is costly.

\begin{figure}[tb!]
\begin{center}
\includegraphics[width=.42\textwidth]{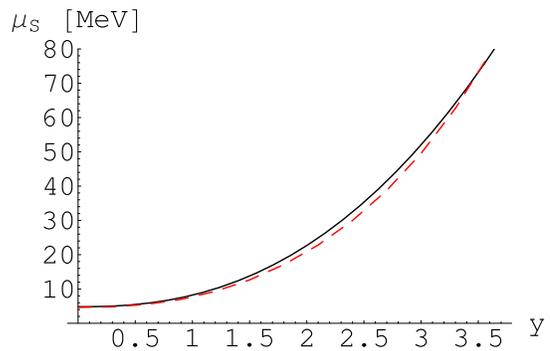}
\end{center}
\vspace{-6mm}
\caption{(Color online) Comparison of the strange chemical 
potential obtained from the fit to the data (solid line) and 
from the condition of zero local strangeness density, 
$\rho_S=0$ (dashed line). \label{fig:sc}}
\end{figure}

There is another important point. In thermal models one may obtain the 
local value of the 
strange chemical potential, $\mu_S$, at a given $\mu_B$ with the condition of 
the vanishing strangeness density, $\rho_S=0$. 
The result is shown in Fig.~\ref{fig:sc}, where we compare the strange chemical 
potential obtained from the fit to the data (solid line) and 
from the condition of zero local strangeness density at 
a given $\mu_B(\alpha_\parallel)$. 
The two curves
turn out to be virtually the same. This shows that the 
net strangeness density in our fireball is, within uncertainties of parameters, 
{\em compatible with 0}. This is not obvious from the outset, 
as the condition of zero strangeness density is not assumed in our fitting procedure.
Although this feature is natural in particle 
production mechanisms, in principle only the total strangeness,
integrated over the whole fireball, must be initially zero. 
Variation of the strangeness density with $\alpha_\parallel$ is admissible, 
but turns out not to occur. 
 
Figure \ref{fig:ratiosy} shows the quality of our fit for the 
parameters of chemical potentials, 
Eq.(\ref{numpar}). 
We show the measured
ratios of $p/{\bar{p}}$, $K^{+}/K^{-}$, and $\pi ^{+}/\pi ^{-}$ as a
function of rapidity $y$ \cite{Bearden:2003fw,Bearden:2004yx} and the
results of fit made with help of the simulation with {\tt THERMINATOR}. 
We note a very reasonable agreement. The error
bars on the model points are statistical errors due to the finite size of the
sample (we use 2500 simulated events in this plot). The flat character
of the $\pi ^{+}/\pi ^{-}$ ratio
indicates that the value of the isospin chemical potential is consistent
with zero at all spatial rapidity values. 

\begin{figure}[tb]
\begin{center}
\includegraphics[width=.45\textwidth]{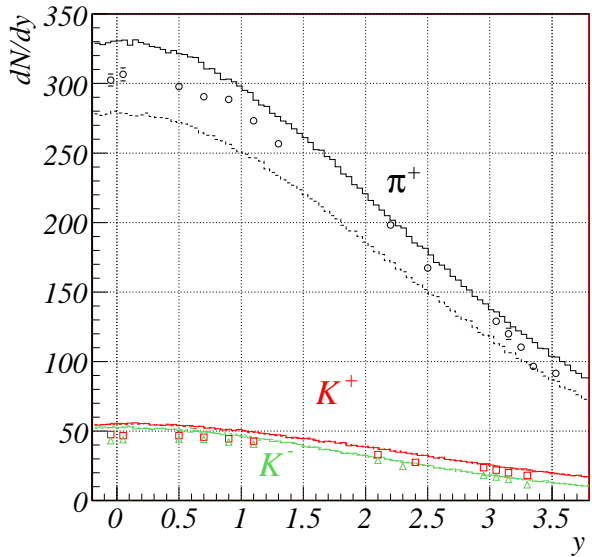}
\end{center}
\vspace{-8mm}
\caption{(Color online)
Rapidity spectra of $\pi^+$, $K^+$, and $K^-$. The data points come
from the BRAHMS collaboration \cite{Bearden:2003fw,Bearden:2004yx} 
(circles - $\pi^+$, squares - $K^+$, triangles - $K^-$),
while the histogram lines show the result of the model simulation with {\tt %
THERMINATOR}. For $\pi^+$ the solid (dashed) line corresponds to the 
full feeding (no feeding) from the weak hyperon decays. 
The model parameters are from Eq.~(\ref{numpar}). 
The experimental pion yields are corrected 
for weak decays as described in \cite{Bearden:2004yx}.
\label{fig:bry}}
\end{figure}

In Fig.~\ref{fig:bry} we show the comparison of obtained 
rapidity spectra of $\pi^+$, $K^+$, and $K^-$ to the experimental data. 
The experimental yields for the pions
are corrected for the feed-down from the weak decays as described in 
Ref.~\cite{Bearden:2004yx}. For that reason
for the case of 
$\pi^+$ we give the model predictions with the full feeding from the weak 
decays (solid line)
and with no feeding from the weak decays at all (dashed line). We note a quite 
good quantitative agreement, with the data 
falling between the two extreme cases. We recall that the behavior 
on rapidity of $dN/dy$ is
controlled by the $\Delta$ parameter of Eq.~(\ref{eq:Delta}). 
The spectra of $\pi^-$ are not shown, since they are practically 
equal to the case of $\pi^+$.
The spectra of $K^+$ and $K^-$ are also quite well reproduced.    

\begin{figure}[tb]
\begin{center}
\subfigure{\includegraphics[width=.45\textwidth]{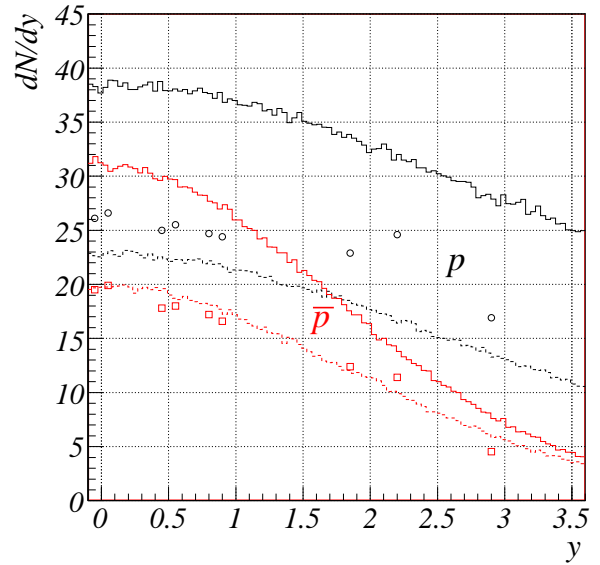}}\\
\vspace{-4mm}
\subfigure{\hspace{-4mm}\includegraphics[width=.47\textwidth]{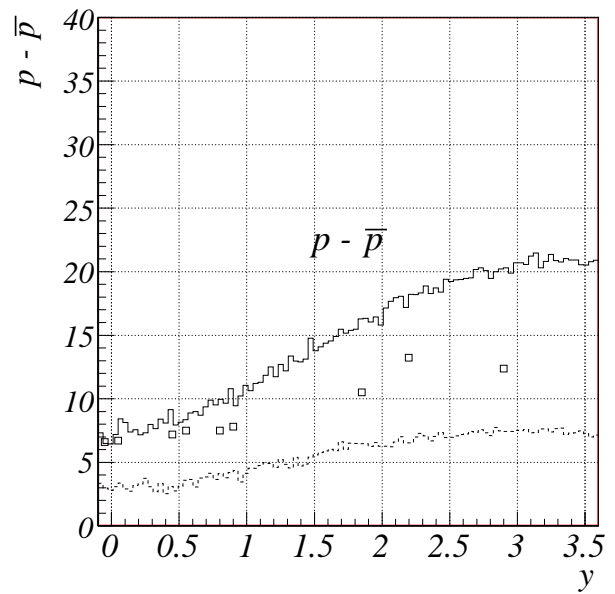}}
\end{center}
\vspace{-8mm}
\caption{(Color online) Top:
the rapidity spectra of $p$ and $\bar p$. Bottom: spectrum of net protons, $p- \bar p$.
The data points come 
from the BRAHMS collaboration \cite{Bearden:2003fw,Bearden:2004yx},
while the solid (dashed) histogram lines show the result of the model simulation with {\tt %
THERMINATOR} with full feeding (no feeding) from the weak hyperon decays. 
Data points should be compared to the model with full feeding (solid lines).
The model parameters are from Eq.~(\ref{numpar}).
\label{fig:bry2}}
\end{figure}

Figure~\ref{fig:bry2} displays the rapidity spectra of protons and antiprotons, as well 
as their difference $p - \bar p$, {\em i.e.} the {\em net} protons. 
Since the $p$ and $\bar p$ data carry no feed-down corrections 
for weak decays \cite{Bearden:2003fw}, one should
compare the solid lines to the data. 
The shape of the $p$ and $\bar p$ spectra is properly 
reproduced, but the model overshoots the data by about 50\%.
This feature occurs at all rapidities, also at mid-rapidity.
The mismatch could be improved by decreasing $T$ by a few percent and redoing the 
whole analysis, but we do 
not take the effort here, holding to the value (\ref{T}) obtained from 
global fits to all RHIC data for the particle yields at midrapidity.
We provide, however, the results of the model calculation with no feeding from the 
hyperon decays, since it provides some measure of the systematic uncertainties
in determining the proton and antiproton yields.
   
Quite remarkably, the 
qualitative growing of the net-proton spectrum with $y$ is obtained. 
This has a simple explanation on the ground of statistical models, since
approximately $p-\bar p \sim {\rm sinh}(\mu_B(y)/T)$. 
Thus at RHIC the proton and antiproton spectra may be qualitatively explained 
solely on the ground of the statistical approach. 

\begin{figure*}[tb]
\begin{center}
\subfigure{\includegraphics[width=.47\textwidth]{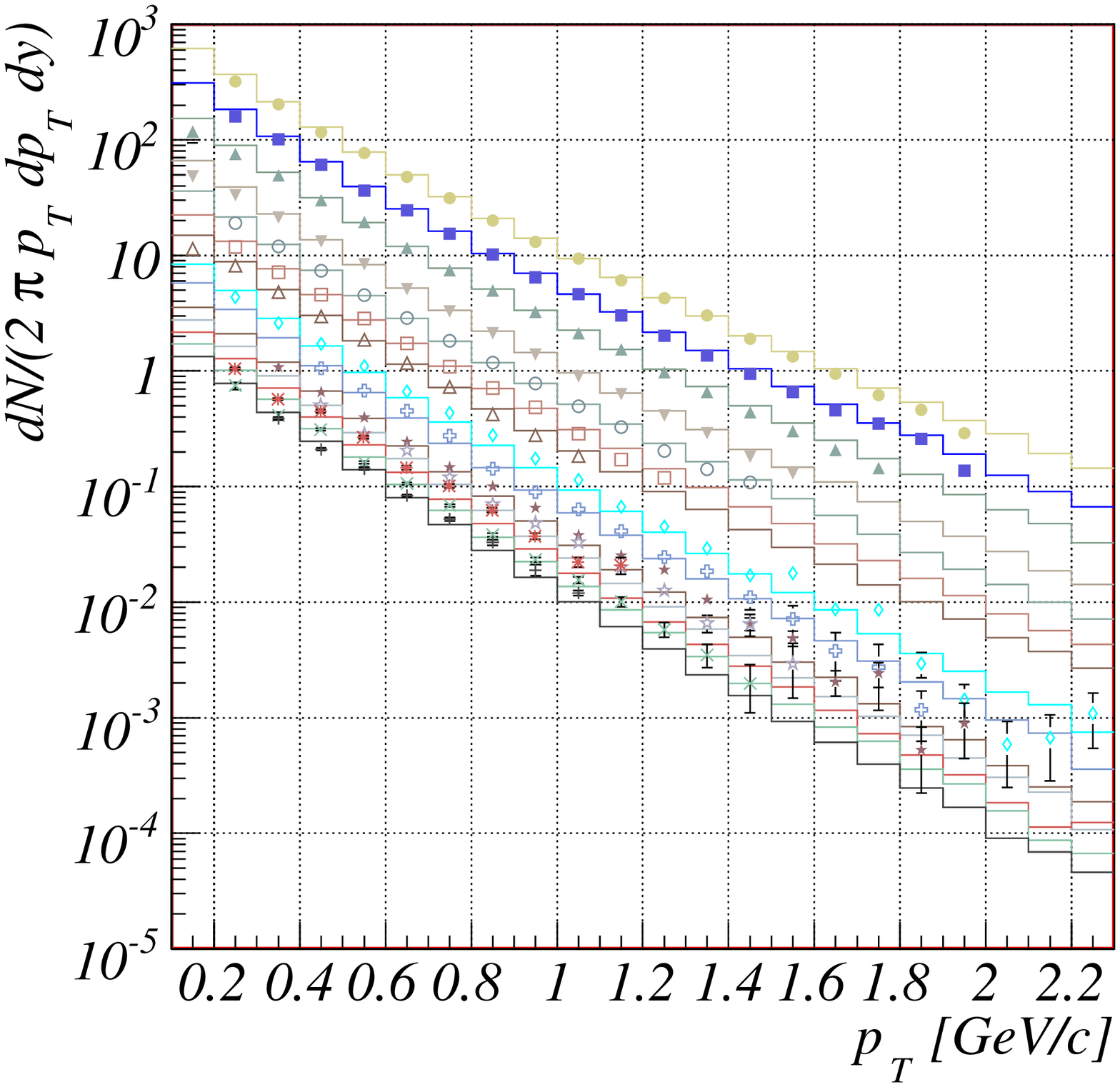}} 
\subfigure{\includegraphics[height=7.93cm,width=.47\textwidth]{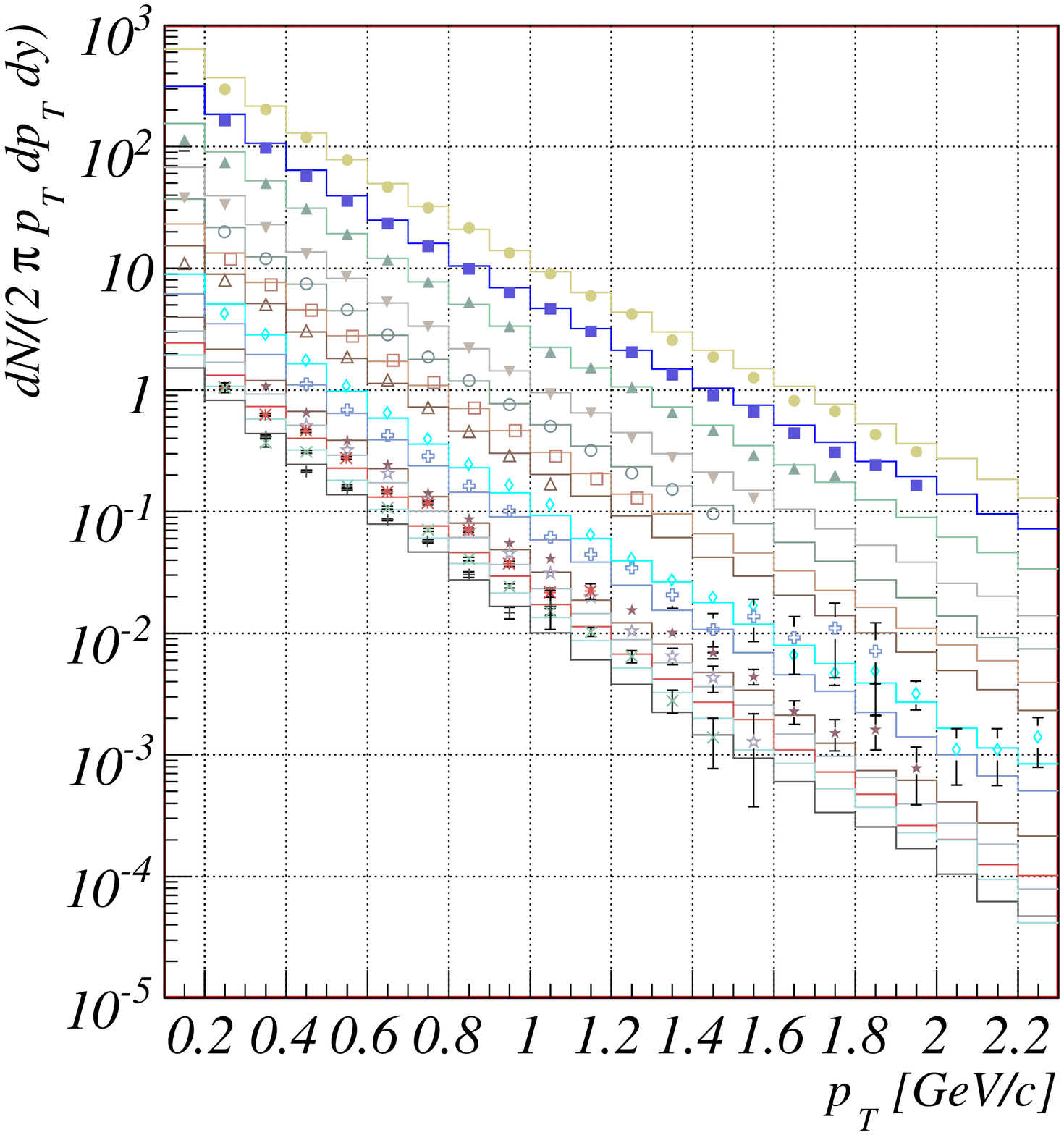}}\\
\vspace{-4mm}
\subfigure{\includegraphics[width=.47\textwidth]{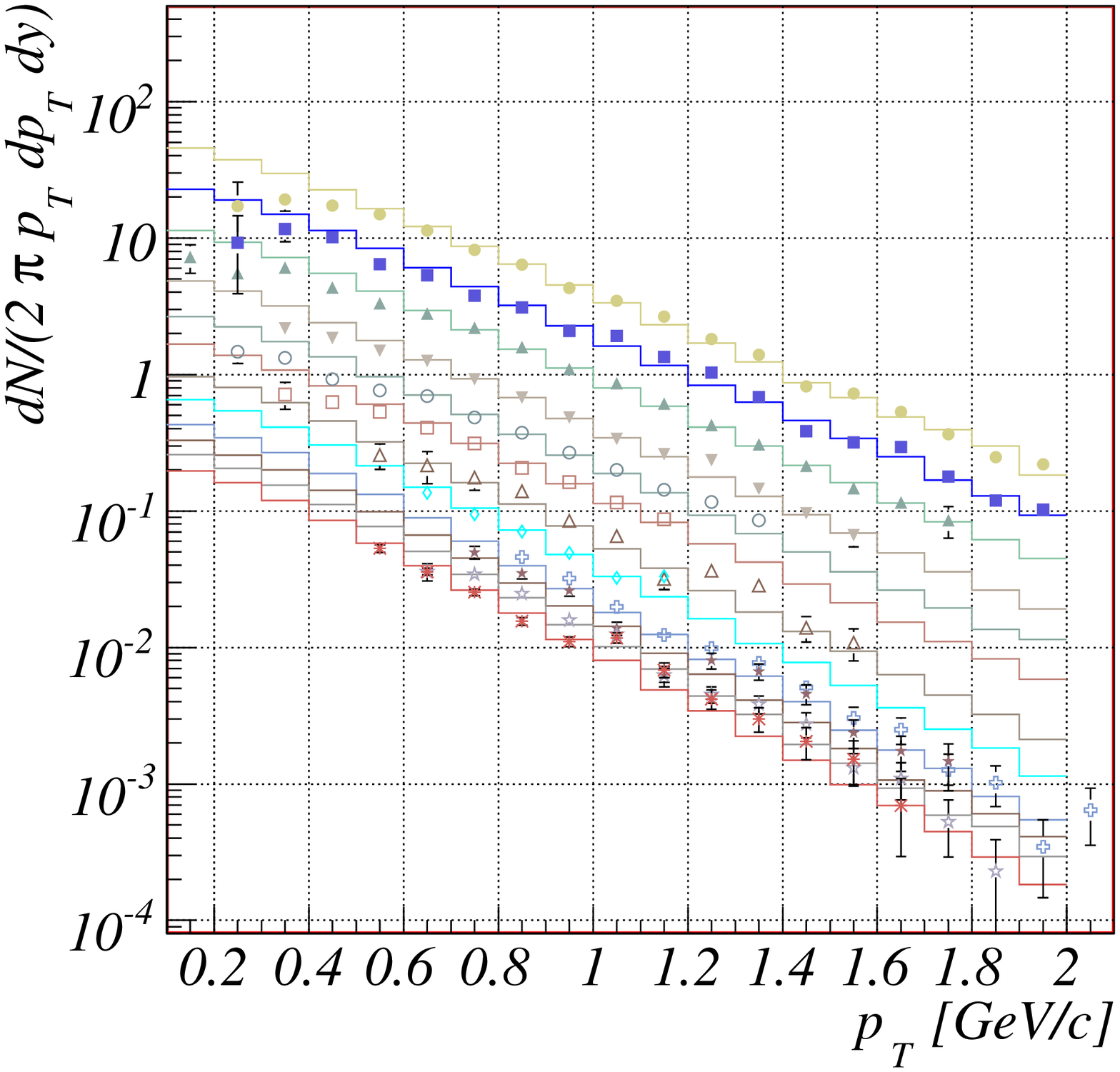}} 
\subfigure{\includegraphics[width=.47\textwidth]{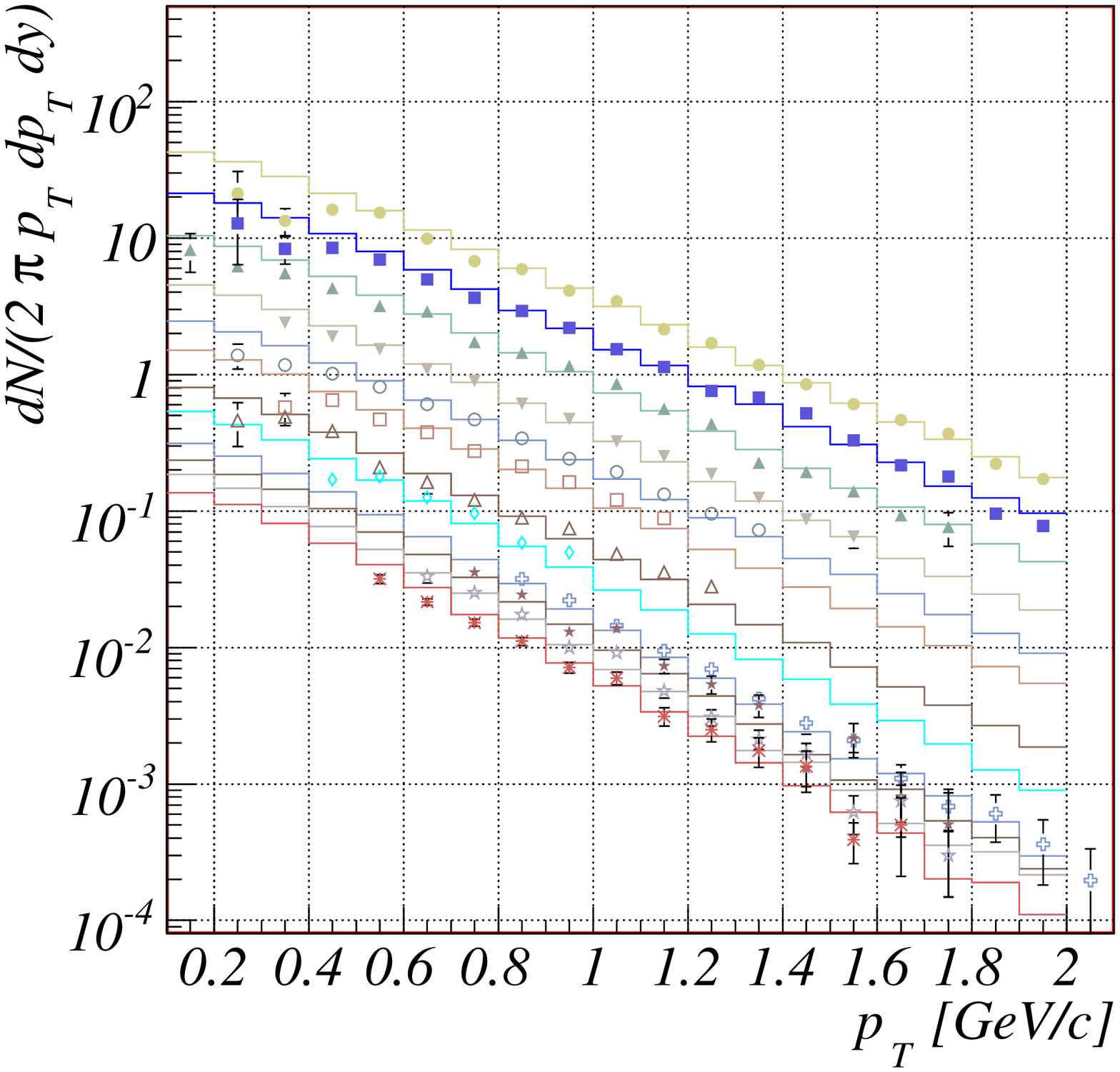}}
\end{center}
\vspace{-8mm}
\caption{(Color online) The $p_{T}$-spectra of $\pi ^{+}$ (top left), 
$\pi ^{-}$ (top right), $K^{+}$
(bottom left), and $K^{-}$ (bottom right) in the subsequent BRAHMS rapidity bins, see the text
for details. The data points come from Ref.~\cite%
{Bearden:2004yx}, while the histogram lines show the result of the
model simulation with {\tt THERMINATOR}.
\label{fig:piKpt}}
\end{figure*}

\begin{figure*}[tb]
\begin{center}
\subfigure{\includegraphics[width=.47\textwidth]{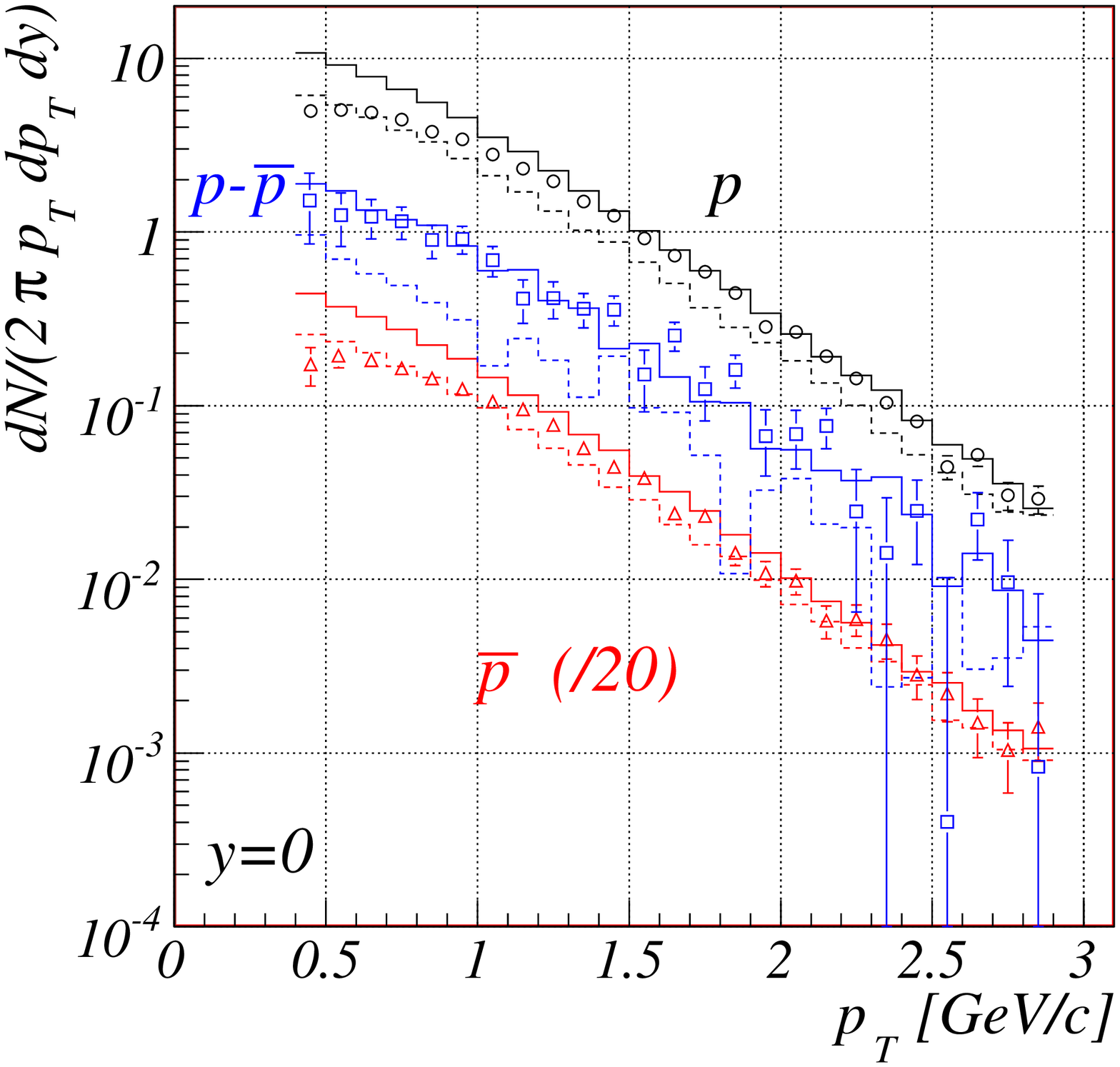}}
\subfigure{\includegraphics[width=.47\textwidth]{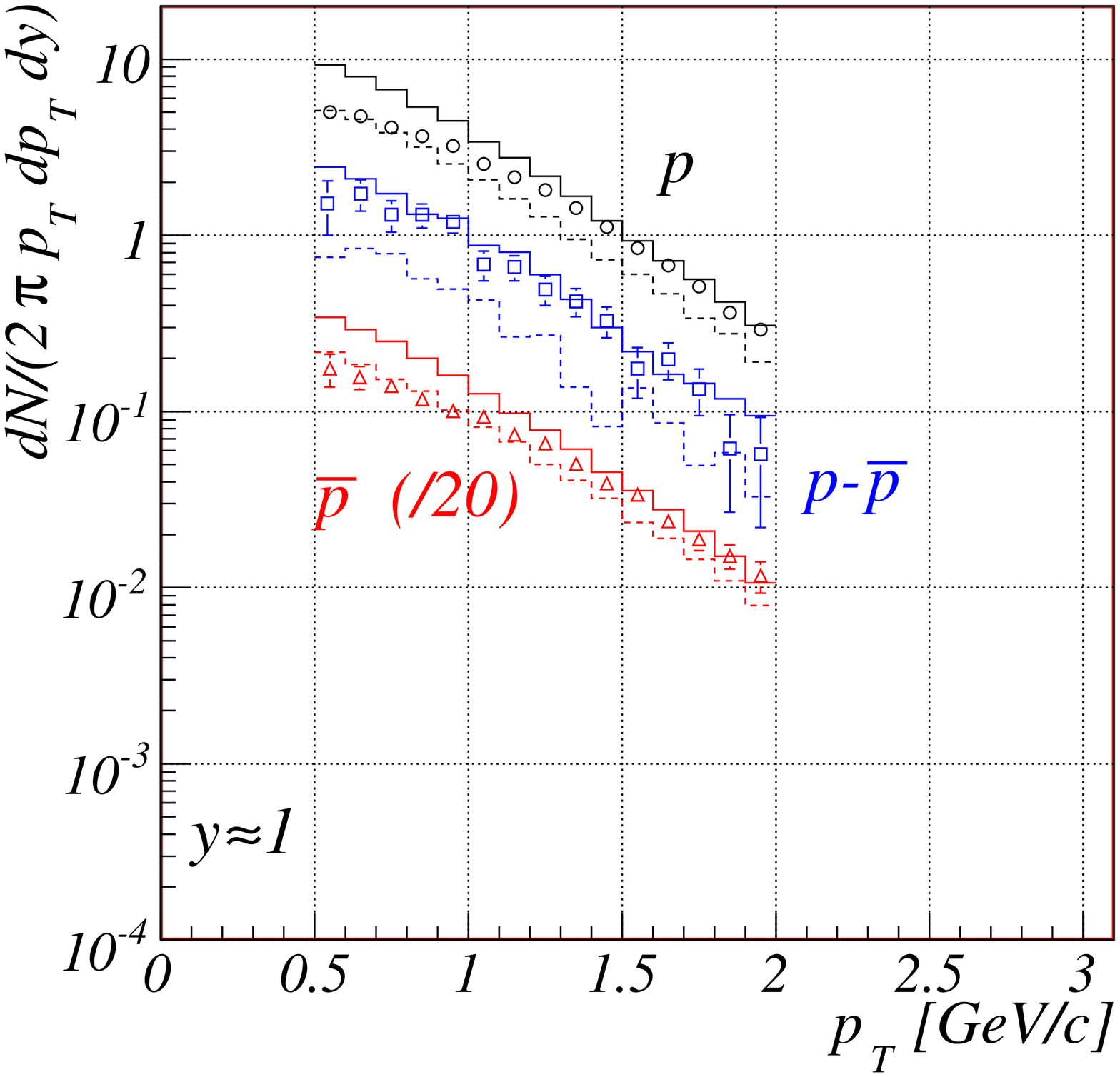}}\\
\vspace{-4mm}
\subfigure{\includegraphics[width=.47\textwidth]{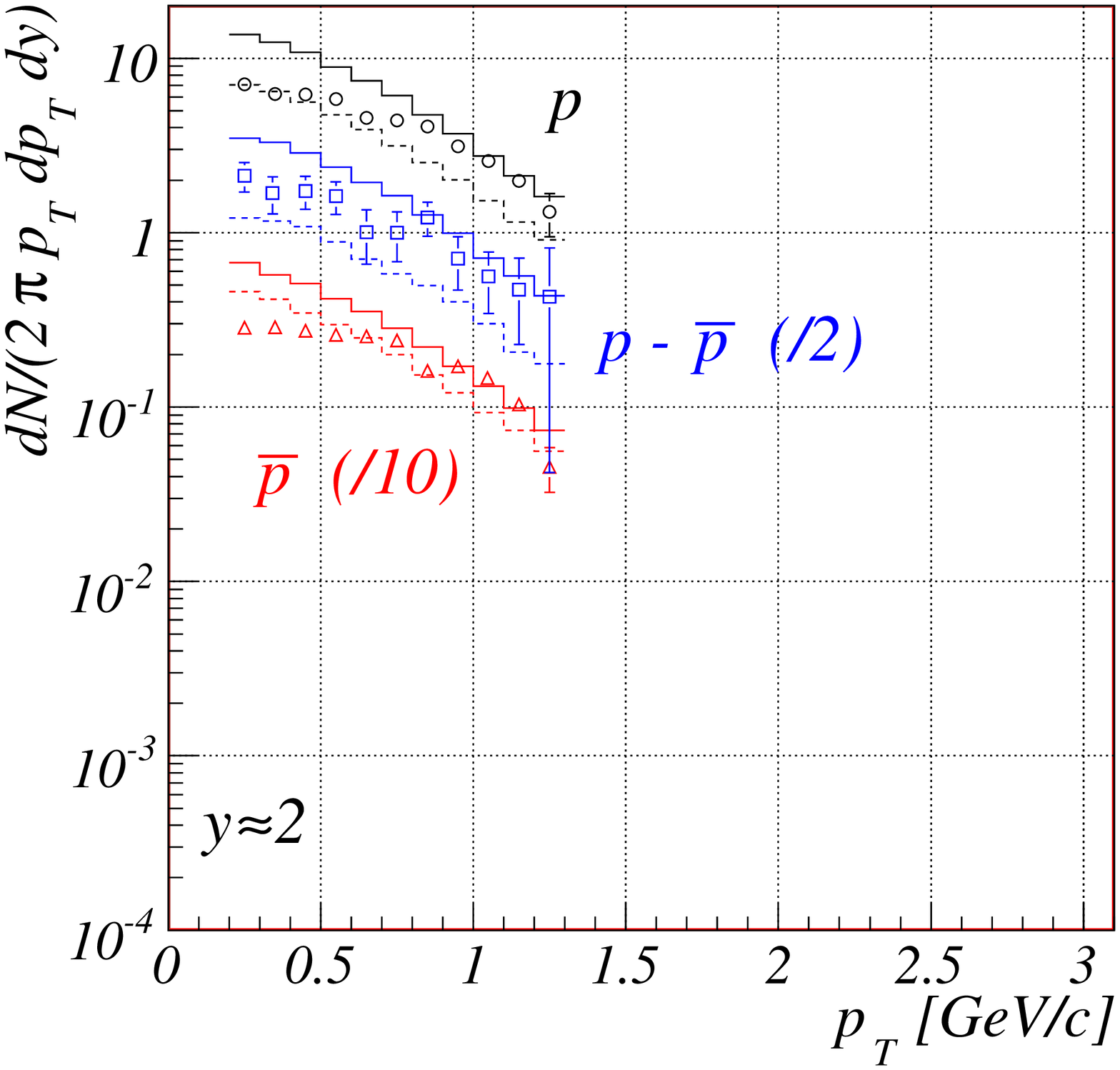}}
\subfigure{\includegraphics[width=.47\textwidth]{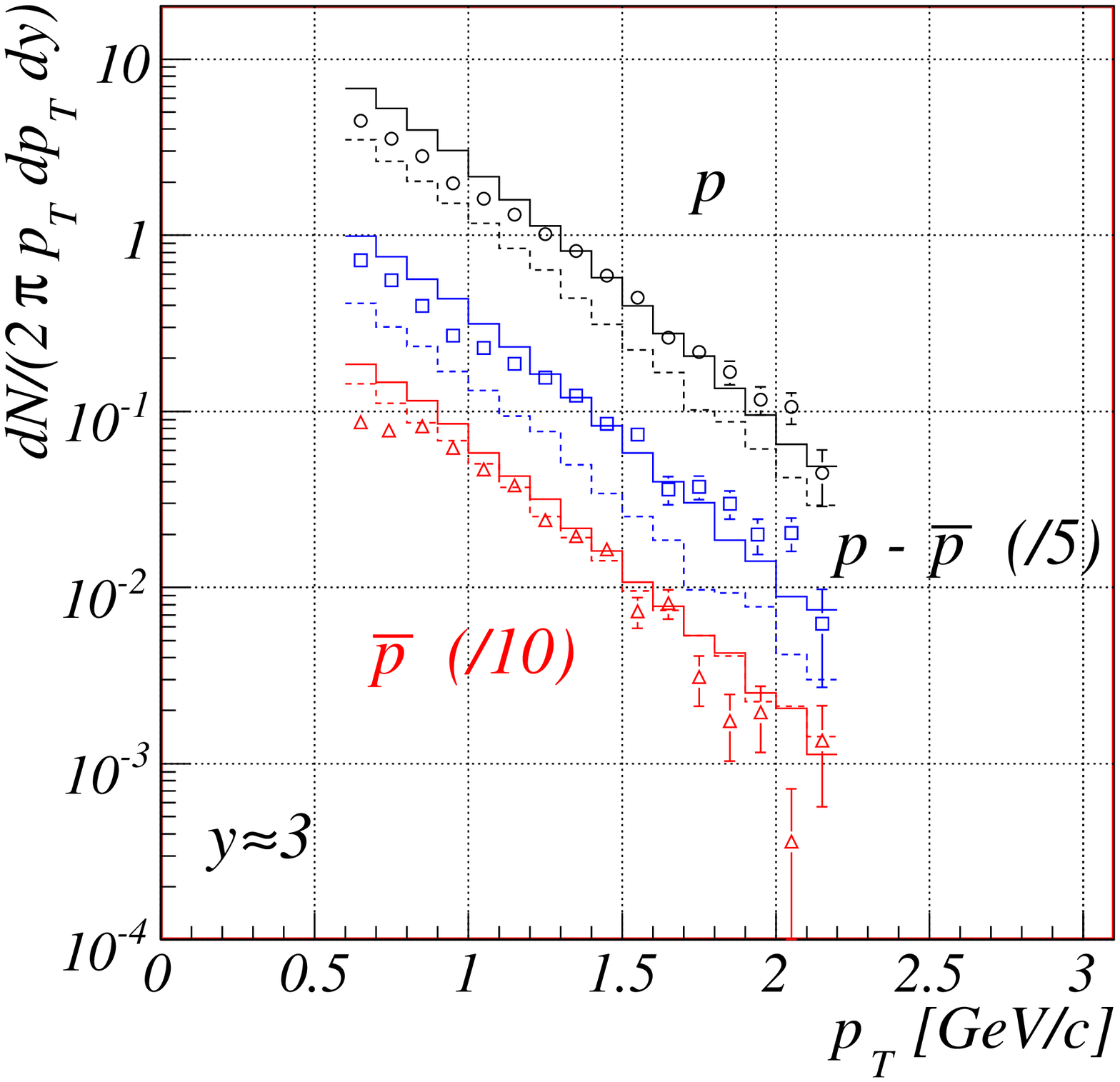}}
\end{center}
\vspace{-8mm}
\caption{(Color online)
The proton, antiproton, and net-proton transverse momentum spectra at $y \simeq 0, 1, 2, 3$. 
The antiproton spectra and the net-proton 
spectra at $y \simeq 3$ have been divided by $10$. The 
data points come from the BRAHMS collaboration \cite{Bearden:2003fw}
and contain no weak-decay corrections.
Solid (dashed) histogram lines show the result of the model simulation with full 
feeding (no feeding) from the weak hyperon decays. Solid lines should be compared to 
the data.
\label{fig:pry}}
\end{figure*}

Figure \ref{fig:piKpt} shows the $p_{T}$-spectra at subsequent rapidity bins
of the BRAHMS experiment. From top to bottom we have for pions 
$y\in [-0.1,0.0]$, $[0.0,0.1]$,$[0.4,0.6]$, $[0.6,0.8]$, $[0.8,1.0]$, $[1.0,1.2]$, $%
[1.2,1.4]$, $[2.1,2.3]$, $[2.4,2.6]$, $[3.0,3.1]$, $[3.1,3.2]$, $[3.2,3.3]$, 
$[3.3,3.4]$, $[3.4,3.66],$ and for kaons $y\in \lbrack -0.1,0.0]$, $%
[0.0,0.1] $,$[0.4,0.6]$, $[0.6,0.8]$, $[0.8,1.0]$, $[1.0,1.2]$, $[2.0,2.2]$, 
$[2.3,2.5] $, $[2.9,3.0]$, $[3.0,3.1]$, $[3.1,3.2]$, and $[3.2,3.4]$. Each lower
curve is subsequently divided by the factor of 2 in order to avoid
overlapping. The solid lines show the model calculation with optimum
parameters (\ref{numpar}). We note that the basic features of the experiment are
reproduced, with the slope increasing with $y$. This can
be explained with a lower transverse flow at larger $y$, as enforced by the
parameterization (\ref{eq:Delta}). The quality of the agreement is similar
in all rapidity bins. In Fig.~\ref{fig:pry} we give the similar study for the 
protons and antiprotons.  The 
data points come from the BRAHMS collaboration \cite{Bearden:2003fw}
and contain no weak-decay corrections, hence the 
solid lines should be compared to the data. Nevertheless, as in Fig.~\ref{fig:bry2}, 
we also present the calculation 
with feed-down from the weak decays switched off, as it provides a measure of 
systematic uncertainties. It should also be kept in mind that these 
uncertainties 
are quite large for the $p_T$-spectra of $p$ and $\bar p$, 
as can be inferred from the comparison 
of results of various experimental collaborations at RHIC (cf. for instance Fig.~12 
of Ref.~\cite{Ullrich:2002tq}). 

\begin{figure*}[tb]
\begin{center}
\subfigure{\includegraphics[width=.45\textwidth]{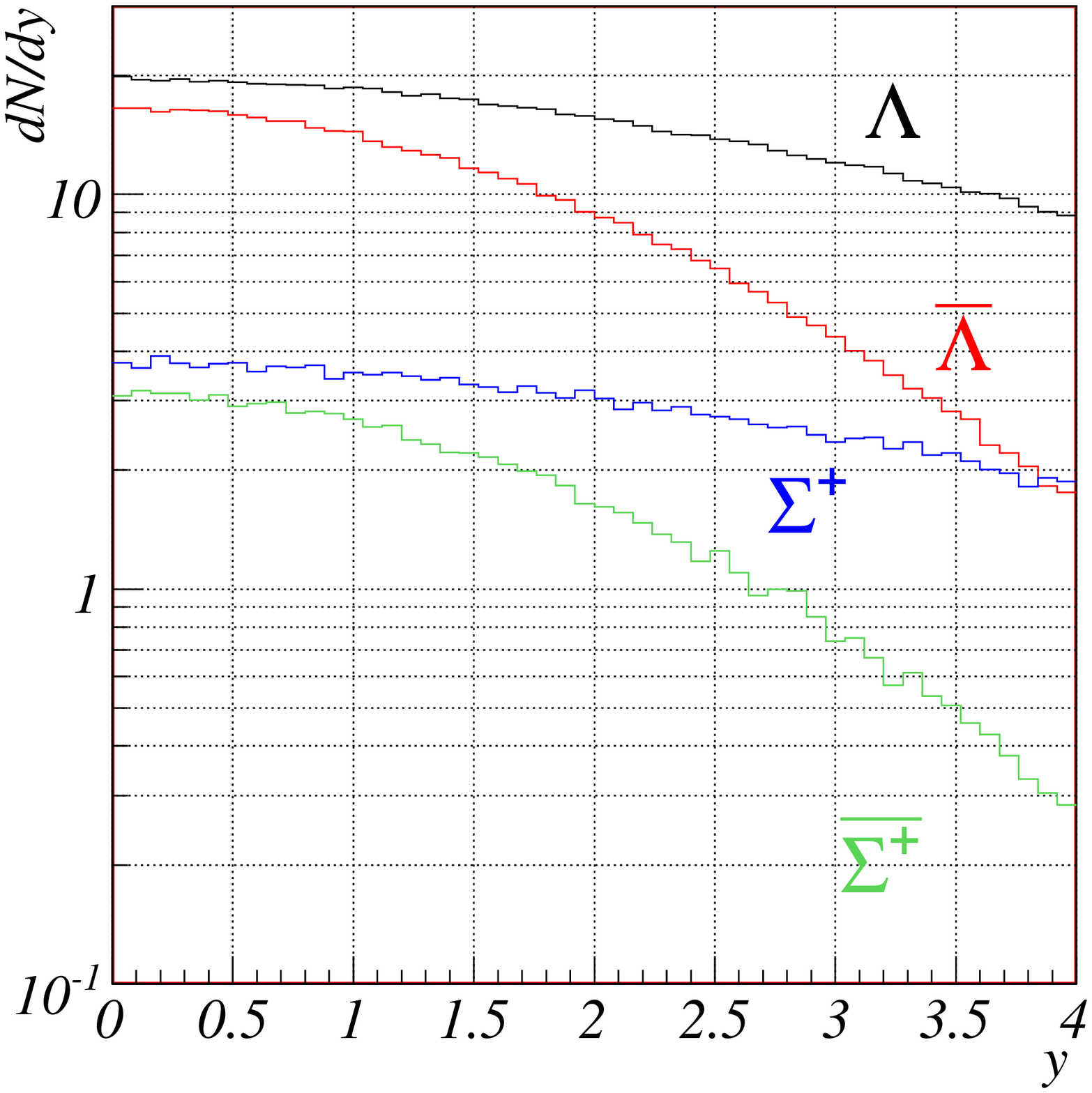}}
\subfigure{\includegraphics[width=.45\textwidth]{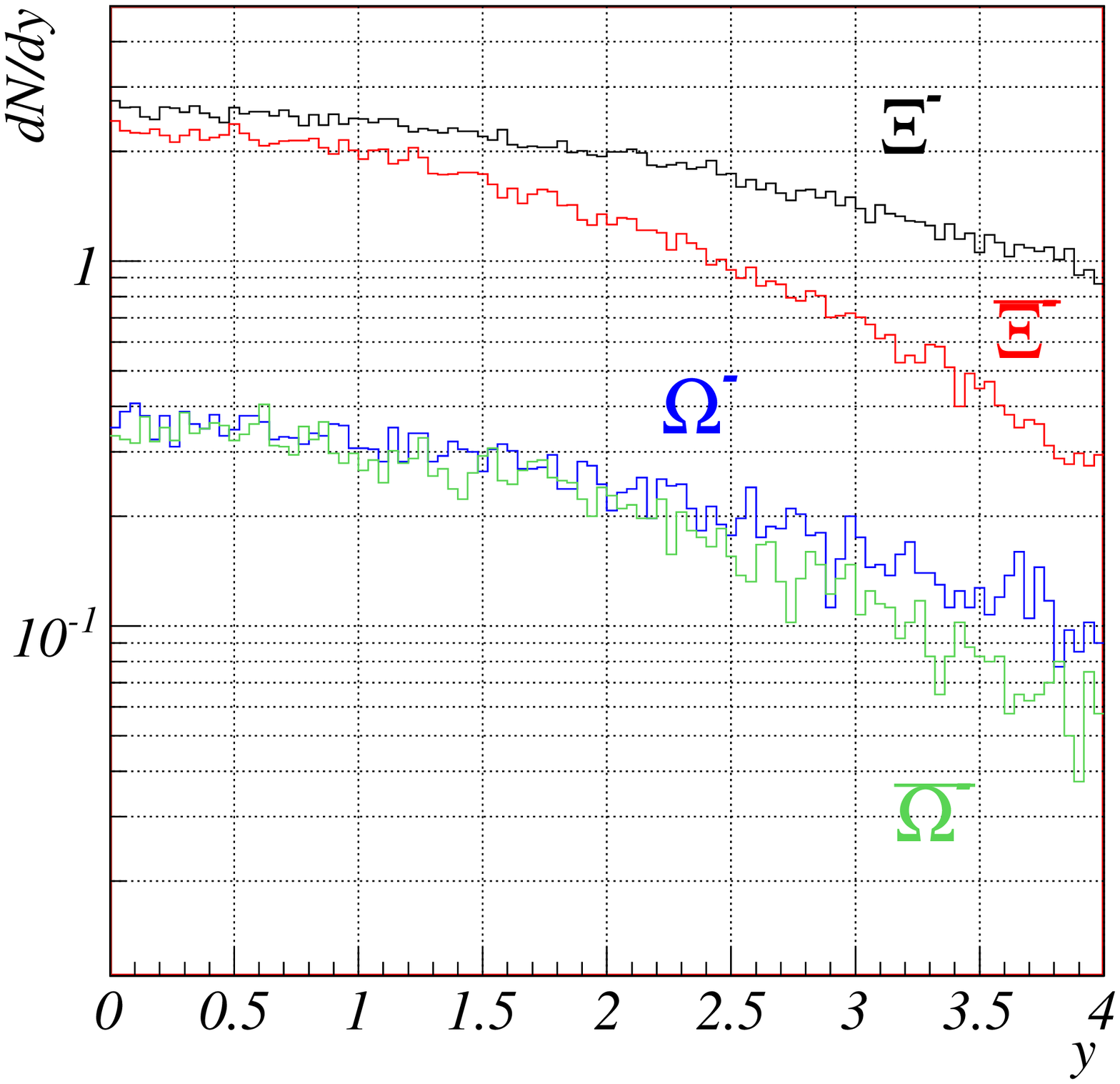}}
\end{center}
\vspace{-8mm}
\caption{(Color online) The model predictions
for the rapidity spectra of hyperons. Top panel, curves from top to bottom: $\Lambda$, $\bar \Lambda$, 
$\Sigma^+$, and $\bar \Sigma^+$. Bottom panel, curves from top to bottom: 
$\Xi^-$,  $\bar \Xi^-$, $\Omega$,  and $\bar \Omega$. 
The model parameters as in Fig.~\ref{fig:ratiosy}. 
\label{fig:hyper}}
\end{figure*}

At this point we have accomplished the goal of fixing the ``fireball topography'':
we have the geometry/flow as well as thermal parameters dependent on the 
variable $\alpha_\parallel$. Next, we may proceed as in the case of the boost-invariant model
used at mid-rapidity, and compute many observables in addition to those already used up to fix 
the model parameters. These observables include one-body observables, such as 
spectra of various particles, including hyperons, mesonic resonances, {\em etc.}, as well
as two-body observables related to correlations: HBT radii, balance functions in rapidity, 
or event-by-event fluctuations. Here we only present a sample prediction for 
rapidity spectra of hyperons, shown in Fig.~\ref{fig:hyper}. 
An interesting feature is the very small splitting of $\Omega$ and $\bar \Omega$, which 
results from the fact that $\mu_B-3\mu_S \simeq 0$, cf. Fig.~\ref{fig:muy}.

\section{Conclusion \label{sect:conclusion}}

The paper contains results of the single-freeze-out thermal model for
rapidity-dependent spectra in relativistic heavy-ion collisions. We have
used {\tt THERMINATOR} to run the simulations and the BRAHMS data for $%
\sqrt{s_{NN}}=200~\mathrm{GeV}$ $Au+Au$ collisions to fix the model
parameters. Such a simulation is necessary when the system is not
boost-invariant. It allows for an exact incorporation of the space-time
dependence of thermal parameters, precise inclusion of resonance decays, as
well as incorporation of experimental cuts. The extension of the original
boost-invariant single-freeze-out model includes a modification of the shape
of the fireball, which here becomes narrower as the magnitude of the spatial
rapidity $\alpha_\parallel$ increases, as well as admits the dependence of
the thermal parameters on $\alpha_\parallel$. As a result of a fit to the
BRAHMS data we have obtained the dependence of the freeze-out chemical
potentials on $\alpha_\parallel$. The freeze-out temperature is taken
constant in the considered range of rapidities. With this extension we are
able to properly describe the double $d^2N/(2\pi p_T dp_T dy)$ spectra from
the experiment. We also make predictions for other particles, in particular
for hyperons.

A code incorporating the elastic collisions neglected in the
single-freeze-out approach could be used as an ``afterburner'' starting from
our freeze-out condition. That way a more accurate collision picture could be
achieved. As we have already mentioned, 
a recent study of Ref.~\cite{Nonaka:2006yn} revealed
that for the mid-rapidity $p_T$-spectra the elastic rescattering is not very
important.

Certainly, the scheme of this paper can be used for other collisions where departures from
the boost invariance are significant, in particular for the rich SPS data. As we
have said, the modeling involves the choice of the parameterization for the
shape of the fireball and the velocity field of flow, where in fact we have quite 
a lot of freedom, as well as the
dependence of the thermal parameters at freeze-out on the space-time position.
Accurate data for numerous observables as functions of the rapidity, 
not only abundances and spectra but
also the correlation data (HBT radii, balance functions), would 
greatly help to constrain
the freedom and acquire insight into the space-time evolution picture of
boost-non-invariant systems formed in relativistic heavy-ion collisions.
Most importantly, the knowledge of the dependence of $R_{\rm side}$ on $y$ 
would put constraints on the shape of the fireball. 

Here are the main results of the paper:

\begin{enumerate}

\item Naive extraction of the baryon and strange chemical 
potentials from ratios of $p/\bar p$ and $K^+/K^-$ works 
surprisingly well, as shown in the comparison to the 
full calculation in Fig.~\ref{fig:muy}.

\item The baryon and strange chemical potentials grow 
with $\alpha_\perp$, reaching at $y \sim 3$ values 
close to those of the highest SPS energies of $\sqrt{s_{NN}}=17~{\rm GeV}$.
This agrees with the recent conclusions of Roehrich \cite{Roehrich}. 

\item At mid-rapidity the values of the chemical potentials are even lower
than derived from the previous thermal fits to the data for $|y|\le 1$, with 
our values taking $\mu_B(0)=19~{\rm MeV}$ and $\mu_S(0)=5~{\rm MeV}$. 
The reason for this effect 
is that the particle with $|y|\le 1$ originate from a region $|\alpha_\parallel|\le
2$, and on the average the effective values of chemical potentials are larger 
compared to the values at the very origin
(cf. Fig.~\ref{fig:muy}).

\item The local strangeness density of the fireball is compatible with zero at all 
values of $\alpha_\parallel$. Although this feature is natural in particle 
production mechanisms, here it has been obtained independently just from 
fitting the chemical potentials to data.

\item The ratio of the baryon to strange chemical potentials varies very weakly with 
rapidity, ranging from $\sim 4$ at midrapidity to $\sim 3.5$ at larger rapidities.

\item The $d^2 N/(2\pi p_\perp dp_\perp dy)$ spectra of pions 
and kaons are well reproduced, supporting our hypothesis for the shape of the fireball 
in the longitudinal direction.

\item The rapidity shape of the spectra of protons and antiprotons 
measured by BRAHMS \cite{Bearden:2003fw}
is described properly, while the model predict too large normalization,
overproducing
these particles by about 50\%. This suggests a lower value of $T$ by a few percent, 
or presence of non-equilibrium factors. We also note that the feature of an 
increasing yield of 
the net protons with rapidity is obtained naturally, 
explaining the shape of the rapidity dependence on purely statistical grounds. 

\end{enumerate}

\begin{acknowledgments}
We thank Wojciech Florkowski for his interest and numerous helpful comments.
This research has been partly supported by the Polish Ministry of Education and
Science, grant 2~P03B~02828.
\end{acknowledgments}

%\bibliography{liter}

\end{document}